\documentclass[prd,preprintnumbers,nofootinbib,superscriptaddress]{revtex4} 

\usepackage{epsfig,graphicx}
\usepackage{dcolumn}
\usepackage{bm}
\usepackage{latexsym}
\usepackage{amsfonts}
\usepackage[usenames,dvipsnames]{color}
\usepackage{enumerate}
\usepackage{amssymb}
\usepackage{amsmath}

\newcommand{\Comment}[1]{{}}
\definecolor{MyDarkBlue}{rgb}{0.15,0.15,0.45}
\usepackage[linktocpage=true]{hyperref}
\hypersetup{
colorlinks=true,
citecolor=MyDarkBlue,
linkcolor=MyDarkBlue,
urlcolor=MyDarkBlue,
pdfsubject={hep-th}
}

\newcommand{\be}{\begin{equation}}
\newcommand{\ee}{\end{equation}}
\newcommand{\bea}{\begin{eqnarray}}
\newcommand{\eea}{\end{eqnarray}}
\newcommand{\beas}{\begin{eqnarray*}}
\newcommand{\eeas}{\end{eqnarray*}}
\newcommand{\nn}{\nonumber}

\begin{document}

\preprint{IPMU13-0071}

\title{Cosmological perturbations in extended massive gravity}

\author{A. Emir G\"umr\"uk\c{c}\"uo\u{g}lu}
\email{emir.gumrukcuoglu@ipmu.jp}
\affiliation{Kavli Institute for the Physics and Mathematics of the Universe, Todai Institutes for Advanced Study, University of Tokyo, 5-1-5 Kashiwanoha, Kashiwa, Chiba 277-8583, Japan}

\author{Kurt Hinterbichler}
\email{khinterbichler@perimeterinstitute.ca}
\affiliation{Perimeter Institute for Theoretical Physics, 31 Caroline St. N, Waterloo, Ontario, Canada, N2L 2Y5}

\author{Chunshan Lin}
\email{chunshan.lin@ipmu.jp}
\affiliation{Kavli Institute for the Physics and Mathematics of the Universe, Todai Institutes for Advanced Study, University of Tokyo, 5-1-5 Kashiwanoha, Kashiwa, Chiba 277-8583, Japan}

\author{Shinji Mukohyama}
\email{shinji.mukohyama@ipmu.jp}
\affiliation{Kavli Institute for the Physics and Mathematics of the Universe, Todai Institutes for Advanced Study, University of Tokyo, 5-1-5 Kashiwanoha, Kashiwa, Chiba 277-8583, Japan}

\author{Mark Trodden}
\email{trodden@physics.upenn.edu}
\affiliation{Center for Particle Cosmology, Department of Physics and Astronomy, University of Pennsylvania, Philadelphia, PA 19104}

\date{\today}

\begin{abstract}
We study cosmological perturbations around self-accelerating solutions to two extensions of nonlinear massive gravity: the quasi-dilaton theory and the mass-varying theory.  We examine stability of the cosmological solutions, and the extent to which the vanishing of the kinetic terms for scalar and vector perturbations of self-accelerating solutions in massive gravity is generic when the theory is extended.  We find that these kinetic terms are in general non-vanishing in both extensions, though there are constraints on the parameters and background evolution from demanding that they have the correct sign.  In particular, the self-accelerating solutions of the quasi-dilaton theory are always unstable to scalar perturbations with wavelength shorter than the Hubble length.
\end{abstract}

\maketitle

\section{Introduction}

Recent years have seen the development of a non-linear theory propagating the five degrees of freedom of a massive graviton  (dRGT theory \cite{deRham:2010ik,deRham:2010kj}, see \cite{Hinterbichler:2011tt} for a review), without the Boulware-Deser ghost \cite{Boulware:1973my,Hassan:2011hr}.  This theory admits self-accelerating solutions \cite{deRham:2010tw,Koyama:2011xz,Nieuwenhuizen:2011sq,Chamseddine:2011bu,D'Amico:2011jj,Gumrukcuoglu:2011ew,Berezhiani:2011mt}, in which the universe is de Sitter without a cosmological constant in the action.  The Hubble scale of these self-accelerating solutions is of order the mass of the graviton.   Having a light graviton is technically natural \cite{ArkaniHamed:2002sp,deRham:2012ew}, so these solutions are of great interest to account for cosmic acceleration in the late-time universe.

Given any non-trivial solution, it is natural to ask how the perturbations around it behave, and in particular whether there are interesting new effects in the propagation of the associated degrees of freedom.  The perturbation theory for the self-accelerating solutions of dRGT has been studied in \cite{Gumrukcuoglu:2011zh,D'Amico:2012pi,Wyman:2012iw,Khosravi:2012rk,Fasiello:2012rw}.  Freedom from the Boulware-Deser ghost means that around any background, at most five degrees of freedom propagate.  Around a homogeneous and isotropic cosmology, these take the form of one transverse-traceless tensor, one transverse vector and one scalar.  Even though the Boulware-Deser ghost is absent, the kinetic terms of these degrees of freedom can potentially have the wrong sign, in which case they are ghosts, signaling that that particular background is unstable.  

In fact, around the self-accelerating solutions of dRGT theory, the scalar and vector degrees of freedom have vanishing kinetic terms~\cite{Gumrukcuoglu:2011zh,DeFelice:2012mx}.  This result could have critical implications for the cosmology, and it is important to understand the extent to which it is a generic result in these types of models.  There are several avenues one might consider.  Quantum mechanically, kinetic terms may be generated by loops.  Determining the sign of such terms, and hence whether the background propagates ghosts, is then a difficult question whose answer depends in general on the details of the matter propagating in the loops.  If we wish to restore the kinetic terms at the classical level, one avenue is to move away from homogeneous and isotropic cosmologies \cite{Gumrukcuoglu:2012aa,DeFelice:2013awa}.  Another is to keep homogeneity and isotropy, but change the theory by adding more degrees of freedom.  In this paper, we take the latter approach.  We study cosmological perturbations in two extensions of dRGT theory: quasi-dilaton massive gravity, and mass-varying massive gravity.
  
Massive gravity admits an extension to a theory with a global scale symmetry through the inclusion of a specific scalar field, dubbed the {\it quasi dilaton} \cite{D'Amico:2012zv}. We examine self-accelerating background solutions to the quasi-dilaton model, and consider the behavior of cosmological perturbations (see also \cite{Haghani:2013eya}).  We find the conditions under which the backgrounds are free of ghosts.  We find that there is always a ghost-like instability in the scalar sector, for fluctuations of physical wavelength shorter than the Hubble radius.  

The other extension we consider is mass-varying massive gravity, the theory obtained by promoting the mass to a scalar field \cite{D'Amico:2011jj,Huang:2012pe}.  We study the behavior of cosmological perturbations in this model, and find the conditions under which the backgrounds are free of ghosts.

\section{Quasi-dilaton theory\label{quasdisec}}

We start with the quasi-dilaton theory, introduced in \cite{D'Amico:2012zv}.
The action governs a dynamical metric $g_{\mu\nu}$ and a scalar field $\sigma$,\footnote{We thank the authors of \cite{D'Amico:2012zv} for pointing out the existence of the $\xi$ term in the most general quasi-dilaton action.}
\begin{equation}
S= \frac{M_p^2}{2}\int\,d^4x\,\left\{\sqrt{-g}\,\left[R[g] - 2\,\Lambda+ 2\,m_g^2 \left({\cal L}_2 +\alpha_3\,{\cal L}_3 +\alpha_4\,{\cal L}_4\right)-\frac{\omega}{M_p^2}\,\partial_\mu\sigma\,\partial^\nu\sigma
\right] + 2\,m_g^2\,\xi\,\sqrt{-\bar{g}}\,e^{4\,\sigma/M_p}\right\}
\,.\label{quasiaction}
\end{equation}
The part of the action which provides the mass to the graviton is
\begin{eqnarray}
 {\cal L}_2 & = & \frac{1}{2}
  \left(\left[{\cal K}\right]^2-\left[{\cal K}^2\right]\right)\,, \nonumber\\
 {\cal L}_3 & = & \frac{1}{6}
  \left(\left[{\cal K}\right]^3-3\left[{\cal K}\right]\left[{\cal K}^2\right]+2\left[{\cal K}^3\right]\right), 
  \nonumber\\
 {\cal L}_4 & = & \frac{1}{24}
  \left(\left[{\cal K}\right]^4-6\left[{\cal K}\right]^2\left[{\cal K}^2\right]+3\left[{\cal K}^2\right]^2
   +8\left[{\cal K}\right]\left[{\cal K}^3\right]-6\left[{\cal K}^4\right]\right)\,,
\label{lag234}
\end{eqnarray}
where square brackets denote a trace. While these expressions are similar in form to the dRGT theory, in the case of the quasi-dilaton theory, the building block tensor ${\cal K}$ is defined as
\begin{equation}
 {\cal K}^\mu _{\ \nu} = \delta^\mu _{\ \nu} 
 - e^{\sigma/M_p}\,\left(\sqrt{g^{-1}\bar g}\right)^{\mu}_{\ \ \nu}\,,
\label{Kdef}
\end{equation}
where $\bar g_{\mu\nu}$ is a non-dynamical fiducial metric.  The theory is invariant under a global dilation of the space-time coordinates, accompanied by a shift of $\sigma$.  This symmetry rules out a non-trivial potential for $\sigma$.

Throughout the analysis, we choose the fiducial metric to be Minkowski, 
\begin{equation}
\bar g_{\mu\nu} dx^\mu\,dx^\nu = -\,dt^2+\,\delta_{ij} dx^i\,dx^j\,.
\end{equation}

\subsection{Background equations of motion}

For the physical background metric, we adopt the flat FRW ansatz
\begin{equation}
g_{\mu\nu} dx^\mu\,dx^\nu = -N(t)^2\,dt^2+a(t)^2\,\delta_{ij} dx^i\,dx^j\, .
\label{flatfrw}
\end{equation}

To obtain the background equations of motion, it is convenient to introduce time reparametrization invariance, so that we may write a mini-superspace action.  We replace the fiducial metric with
\begin{equation}
\bar{g}_{\mu\nu} dx^\mu\,dx^\nu = -\,f'(t)^2 dt^2+\,\delta_{ij} dx^i\,dx^j\, ,
\end{equation}
where $f(t)$ is the St\"uckelberg scalar \cite{ArkaniHamed:2002sp}, and unitary gauge corresponds to the choice $f(t)=t$.

The mini-superspace action is
\bea \frac{S}{V} &=&\int dt \bigg\{M_p^2 \left[-3{a\dot a^2\over N}-\Lambda a^3 N\right]+{\omega a^3\over 2N}\dot\sigma^2+M_p^2m_g^2\bigg[N a^3(X-1)\left(3 (X-2 ) - (X-4 ) (X-1 ) \alpha_3 - (X-1 
    )^2 \alpha_4\right) \nn\\  
    && \qquad\qquad\qquad\qquad\qquad\qquad\qquad\qquad\qquad+  f' a^4 X \, \left[
(X-1)\left(3  - 3 (X-1 )  \alpha_3 + (X-1 )^2  \alpha_4\right)
+\xi\,X^3\right]
\bigg]\bigg\},
\eea
where $V$ is the comoving volume and we have defined
\begin{equation}
X \equiv \frac{e^{\sigma/M_p}}{a}\,.
\end{equation}
In addition, to simplify expressions later on, we define
\be H\equiv {\dot a \over Na},\ \ \ r\equiv \frac{a}{N}.\ee

Varying with respect to $f$ and then choosing unitary gauge, we obtain a constraint equation:

\begin{equation}
\left.\frac{\delta\,S}{\delta\,f}\right|_{f=t} = m_g^2\,M_p^2\,\frac{d}{dt}
\Big\{
a^4\,X\Big[(1-X)\,
\left[
3-3 (X-1)\alpha_3 + (X-1)^2\alpha_4
\right] - \xi\,X^3\Big]
\Big\}=0\,.
\label{eqstuck}
\end{equation}

The Friedmann equation is obtained by varying with respect to the lapse $N$, 
\begin{equation}
{1\over M_p^2 a^3}\left.\frac{\delta\,S}{\delta\,N}\right|_{f=t} =3\,H^2 -\Lambda -m_g^2 (X-1)\left[
-3 (X-2)+ (X-4) (X-1)\alpha_3+ (X-1)^2 \alpha_4
\right] -
\frac{\omega}{2}
\left(H + \frac{\dot{X}}{X\,N}\right)^2=0\, ,
\label{eqn}
\end{equation}
and varying with respect to the scale factor, $a$, yields the acceleration equation. This may be combined in a linear combination with \eqref{eqn}) to yield the simpler equation
\bea
{1\over 6 M_p^2 a^2 N}\left.\frac{\delta\,S}{\delta\,a}\right|_{f=t}&+&{1\over 2 M_p^2 a^3}\left.\frac{\delta\,S}{\delta\,N}\right|_{f=t} =
\nn\\ && \frac{\dot{H}}{N}+\frac{\omega}{2}\left(H+\frac{\dot{X}}{N\,X}\right)^2+ \frac{m_g^2}{2}\,(1-r)\,X\,\left(
3-2 X+ (X-3) (X-1)\alpha_3+ (X-1)^2\alpha_4\right)=0.\nn\\
\label{eqa}
\eea

Finally, the equation of motion for $\sigma$ is
\begin{eqnarray}
&&-{X\over M_p\omega a^3 N}\left.\frac{\delta\,S}{\delta\,\sigma}\right|_{f=t} = \frac{1}{N}\,\frac{d}{dt}\,\left(\frac{\dot{X}}{N}\right)+3\,H\,X\,\left(H+\frac{\dot{X}}{N\,X}\right)+X\,\left[\frac{\dot{H}}{N}-\left(\frac{\dot{X}}{N\,X}\right)^2\right]
\nonumber\\
&&\quad+ \frac{m_g^2\,X^2}{\omega}\,\Bigg[
3r (1-2 X)-6 X+9+3  (X-1) (r (3 X-1)+X-3)\alpha_3- (X-1)^2 (r (4 X-1)-3)\alpha_4-4\,r\,X^3\,\xi
\Bigg]=0. \nn\\
\label{eqs}
\end{eqnarray}

Since time reparametrization invariance was introduced with the St\"uckelberg field $f$, there is a Bianchi identity which relates the four equations,
\be  
{\delta {S}\over \delta \sigma}\dot \sigma+ {\delta {S}\over \delta f}\dot f-N{d\over dt}{\delta {S}\over \delta N}+\dot a{\delta {S}\over \delta a}=0 \ .
\ee
Therefore one equation is redundant with the others and may be dropped.  In discussing solutions we will generally choose to drop the acceleration equation (\ref{eqa}), although we will use it to simplify expressions for the perturbations.

\subsection{Self-accelerating background solutions}

We now discuss solutions, starting with the St\"uckelberg constraint (\ref{eqstuck}). Integrating this equation gives
\begin{equation}
X\,\Big\{(1-X)\,
\left[
3-3 (X-1)\alpha_3 + (X-1)^2\alpha_4
\right]-\xi\,X^3\Big\}
=\frac{1}{a^4}\times {\rm constant}\,.
\label{eq:qmg-stuck}
\end{equation}
In an expanding universe, the right hand side of the above equation decays as $a^{-4}$. Thus after a sufficiently long time, $X$ saturates to a constant value $X_{\rm SA}$, corresponding to a zero of the left hand side of (\ref{eq:qmg-stuck}).
These constant $X$ solutions lead to an effective energy density which acts like a cosmological constant. As pointed out in \cite{D'Amico:2012zv}, there are four such solutions for which $X$ is constant. 
Of these, $X=0$ implies $\sigma\to -\infty$, and as in \cite{D'Amico:2012zv}, we drop this solution to avoid strong coupling.\footnote{As we discuss at the end of Appendix~\ref{qmg-stability}, the remaining solutions also lead to strong coupling in the vector and scalar sectors, when $\xi=0$ and the parameters $\alpha_3$ and $\alpha_4$ are such that $X\simeq0$.} 
What remains are the three solutions to the cubic equation
\footnote{For the choice $\xi=0$, the system simplifies as one of the solutions to Eq.(\ref{cubic}) becomes $X_{\rm SA}=1$, while the remaining two are
\begin{equation}
X_\pm\equiv X_{\rm SA}\Big\vert_{\xi=0} =
\frac{3\,\alpha_3+2\,\alpha_4 \pm \sqrt{9\,\alpha_3^2 -12\,\alpha_4}}{2\,\alpha_4}\,.
\label{xpm}
\end{equation}
In this special setting, the solution $X=1$ is uninteresting: the effective cosmological constant from the mass term is zero and in the present scenario, the background becomes equivalent to a de Sitter universe driven by a (bare) cosmological constant $6\,\Lambda/(6-\omega)$. However, in the presence of matter fields and no bare cosmological constant, this solution asymptotically approaches a Minkowski background and is unstable \cite{D'Amico:2012zv}.}
\begin{equation}
(1-X)\,
\left[
3-3 (X-1)\alpha_3 + (X-1)^2\alpha_4
\right]-\xi\,X^3\Big\vert_{X = X_{\rm SA}}=0\,.
\label{cubic}
\end{equation}
For these solutions, we write the Friedmann equation (\ref{eqn}) as
\begin{equation}
\left(3-\frac{\omega}{2}\right)H^2= \Lambda+ \Lambda_{\rm SA}\,,
\label{friedmann}
\end{equation}
where the effective cosmological constant from the mass term is
\begin{equation}
\Lambda_{\rm SA} \equiv m_g^2\,(X_{\rm SA}-1)\left[
-3 X_{\rm SA}+6+ (X_{\rm SA}-4) (X_{\rm SA}-1)\alpha_3+ (X_{\rm SA}-1)^2\alpha_4
\right]\,.
\end{equation}
The Friedmann equation (\ref{friedmann}) also provides a condition on the parameter $\omega$; on the self-accelerating solutions, one needs to have $\omega<6$ in order to keep the left hand side of the Friedmann equation \eqref{friedmann} positive.  This ensures that when ordinary matter is added to the right hand side, we will have standard cosmology during matter domination.  (Although we do not include matter fields, the sign of the matter energy density can be determined by replacing the bare $\Lambda$ with $\rho/M_p^2$.)

Finally, on the self-accelerating solutions, with constant $H$ specified by (\ref{friedmann}), the equation of motion for
$\sigma$ fixes the ratio $r=a/N$. From Eq.(\ref{eqs}), we obtain
\begin{equation}
r_{\rm SA}=1+ \frac{\omega \,H^2\,(X_{\rm SA}-1)}{m_g^2\,X_{\rm SA}^2\,\left[\alpha_3(X_{\rm SA}-1)^2-2(X_{\rm SA}-1)-\xi\,X_{\rm SA}^2\right]}\,.
\end{equation}
Here, to simplify the expression we have used the St\"uckelberg equation (\ref{eq:qmg-stuck}) to eliminate $\alpha_4$.

\subsection{Perturbations\label{pertsub}}

To find the action for quadratic perturbations, we expand the physical metric in small fluctuations $\delta g_{\mu\nu}$ around a solution $g_{\mu\nu}^{(0)}$,
\be 
g_{\mu\nu}=g_{\mu\nu}^{(0)}+\delta g_{\mu\nu}\ ,
\ee
and keep terms to quadratic order in $\delta g_{\mu\nu}$.

We break the perturbations into standard scalar, transverse vector and transverse-traceless tensor parts,
\bea 
&&\delta g_{00} = -2\,N^2\,\Phi\,, \\
&&\delta g_{0i} = N\,a\,\left(B_i^T+\partial_i B\right)\,, \\
&&\delta g_{ij} = a^2\,\left[h_{ij}^{TT}+ {1\over 2}(\partial_i E^T_j+\partial_j E^T_i)+
2\,\delta_{ij}\,\Psi +\left(\partial_i\partial_j - \frac{1}{3}\delta_{ij}\,\partial_l\partial^l\right)E\right]\,,
\eea
where 
\be 
\partial^ih_{ij}^{TT} = h^{TT\;i}_i=0,\ \ \ \partial^iB^T_i=0,\ \ \ \partial^iE^T_i=0\ .
\ee
We then introduce the perturbation of the scalar via
\begin{equation}
\sigma = \sigma^{(0)} +M_p\,\delta\sigma \ .
\label{eq:qmg-dilaton0}
\end{equation}

We perform the entire analysis in unitary gauge, so that there are no issues of gauge invariance to worry about, and no need to form gauge invariant combinations.  We write the actions expanded in Fourier plane waves, i.e. $\vec \nabla^2\rightarrow -k^2$, $d^3x\rightarrow d^3k$.  Raising and lowering
of the spatial indices on perturbations is always carried out by $\delta^{ij}$ and
$\delta_{ij}$. 

\subsection{Tensor perturbations}

We begin by considering tensor perturbations around the background (\ref{flatfrw}),
\begin{equation}
\delta g_{ij}=a^2 h_{ij}^{TT}\,,
\end{equation}
where $\partial^ih_{ij}^{TT} = h^{TT\;i}_i=0$.
The tensor quadratic action reads 
\begin{equation}
S= \frac{M_p^2}{8}\,\int d^3k\,a^3\,N\,dt\,\left(\frac{1}{N^2}|\dot{h}^{TT}_{ij}|^2-\left( \frac{k^2}{a^2} +M_{GW}^{2}\right)|h^{TT}_{ij}|^2\right)\,,
\end{equation}
where the mass of the tensor modes is given by 
\begin{equation}
M_{GW}^2 \equiv  \frac{m^2_g\,(r-1)\,X_{\rm SA}^3}{X_{\rm SA}-1}\,\left(1+\frac{\xi\,X_{\rm SA}}{X_{\rm SA}-1}\right) +H^2\,\omega\,\left(\frac{r}{r-1}+\frac{2}{X_{\rm SA}-1}\right)\,.
\label{mgwqmgdef}
\end{equation}

To obtain this, we have first used the background acceleration equation to eliminate any terms with $\ddot a$.  Then we have used the self-accelerating branch of \eqref{eq:qmg-stuck} (at late times when the right hand side is zero), the Friedman equation \eqref{eqn} evaluated on the self-accelerating solution (i.e. $\dot X=0$), and the $\sigma$ equation \eqref{eqs} evaluated on the self-accelerating solution (i.e. $\dot X=\dot H=0$), to eliminate $\Lambda$, $\alpha_3$ and $\alpha_4$.

The tensor mode always has correct sign kinetic and gradient terms.  However, it will be tachyonic if the mass term is negative: $M_{GW}^2<0$.  The stability of long wavelength gravitational waves is thus ensured by the condition $M_{GW}^2>0$.  Nevertheless, even if this condition is violated, the tachyonic mass is generically of order Hubble, so the instability would take the age of the universe to develop.

\subsection{Vector perturbations}
\label{qmg-vector}

We next turn to vector perturbations,
\begin{equation}
\delta g_{0i} = N\,a\,B_i^T\,,
\qquad
\delta g_{ij} = \frac{a^2}{2}\,(\partial_i E^T_j+\partial_j E^T_i)\,,
\end{equation}
where $\partial^iB^T_i=\partial^iE^T_i=0$.  The field
$B^T_i$ enters the action without time derivatives, so we may eliminate it as an auxiliary field using its own equation of motion (again we are using the equations of the background self-accelerating solution to eliminate $\Lambda$, $\alpha_3$ and $\alpha_4$)
\begin{equation}
B^T_i = \frac{k^2\,a\,(r^2-1)}{4\,\omega\,a^2\,H^2 + 2\,k^2\,(r^2-1)}\,\frac{\dot{E}^T_i}{N}\,.
\end{equation}
Once this is inserted back into the action, what remains is a system of a single propagating vector,
\begin{equation}
S= \frac{M_p^2}{8}\,\int d^3k\,a^3\,N\,dt\,\left(\frac{{\cal T}_V }{N^2}|\dot{E}^T_{i}|^2-\frac{k^2\,M_{GW}^2}{2}|E^T_{i}|^2\right)\,,
\label{qmg-vec-act1}
\end{equation}
where
\begin{equation}
{\cal T}_V \equiv \frac{k^2}{2}\,\left(1+ \frac{k^2(r^2-1)}{2a^2\,H^2\,\omega}\right)^{-1}\,,
\label{kinvec}
\end{equation}
and $M_{GW}$ is the mass of the tensor modes as in~(\ref{mgwqmgdef}).

From Eq.(\ref{kinvec}), we see that for $(r^2-1)/\omega <0$, there exists a
critical momentum scale, $k_{c}=aH{\sqrt{2\omega\over 1-r^2}}$, above which the vector becomes a ghost. In the case $(r^2-1)/\omega \geq 0$, the
kinetic terms of vector always has correct sign and thus
there is no such critical momentum scale.  In the first case, stability of the system
 requires that the physical critical momentum scale, $k_c/a$, be above the ultraviolet cutoff scale of the effective field theory, $\Lambda_{UV}$, i.e.
\begin{equation}
\Lambda_{UV}^2\lesssim {2H^2\omega\over 1-r^2} \,,\qquad
{\rm if }\; (r^2-1)/\omega <0
\ .
\label{ghostvec}
\end{equation}

To determine whether the vector modes suffer from other instabilities, we define canonically normalized fields,
\begin{equation}
{\cal E}_i^T \equiv \frac{M_p}{2}\,{\cal T}_V\,E_i^T\,,
\end{equation}
in terms of which, the action (\ref{qmg-vec-act1}) reads
\begin{equation}
S= \frac{1}{2}\,\int d^3k\,a^3\,N\,dt\,\left(\frac{1}{N^2}|\dot{\cal E}^{T}_{i}|^2-\omega_V^2|{\cal E}^{T}_{i}|^2\right)\,,
\end{equation}
where the dispersion relation of the modes is given by
\begin{equation}
\omega^2_V = (1+q^2)\,M_{GW}^2- \frac{H^2\,q^2(1+4\,q^2)}{(1+q^2)^2}\,,
\label{qmg-omegav}
\end{equation}
and we have defined the dimensionless quantity
\begin{equation}
q^2 \equiv \frac{k^2}{a^2}\,\frac{r^2-1}{2\,H^2\,\omega}\,.
\end{equation}
The second term in the dispersion relation (\ref{qmg-omegav}), which
originates from the time derivatives of ${\cal T}_V$, is always of order
${\cal O}(H^2)$, provided that $q^2>0$. Therefore, in this regime, this
term does not introduce instabilities faster than the Hubble expansion
rate. Moreover, if $q^2<0$, the no-ghost condition (\ref{ghostvec})
imposes $\vert q^2\vert \lesssim (k^2/a^2)/\Lambda_{UV}^2$. 
Thus, for any physical momenta sufficiently lower than the cutoff
scale of the effective theory, the second term in (\ref{qmg-omegav})
does not lead to any visible instability, i.e. the growth rate of any
instability (if any exist) is at most of the cosmological scale.

The vector modes may potentially suffer from a gradient instability
arising from the first term in (\ref{qmg-omegav}), if $M_{GW}^2<0$ and
$q^2>0$. The growth rate of this instability can be made lower than or
at most of the order of the cosmological scale for all physical momenta below the UV
cut-off $\Lambda_{UV}$, provided that 
\begin{equation}
\Lambda_{UV}^2 \lesssim \frac{2\,H^2\,\omega}{r^2-1}
\,,\qquad
{\rm if }\; (r^2-1)/\omega >0\;\; {\rm and} \;\; M_{GW}^2 <0\,.
\end{equation}

\subsection{Scalar perturbations} 
\label{qdmgscalar}
Finally we consider the action quadratic in scalar perturbations,
\begin{equation}
\delta g_{00} = -2\,N^2\,\Phi\,,
\qquad
\delta g_{0i} = N\,a\,\partial_i B\,,
\qquad
\delta g_{ij} = a^2\,\left[
2\,\delta_{ij}\,\Psi +\left(\partial_i\partial_j - {1\over 3}{\delta_{ij}}\,\partial_l\partial^l\right)E\right]\,,
\label{eq:qmg-decomp}
\end{equation}
\begin{equation}
\sigma = \sigma^{(0)} +M_p\,\delta\sigma \ .
\label{eq:qmg-dilaton}
\end{equation}

\begin{figure}
\includegraphics[width=7.5cm]{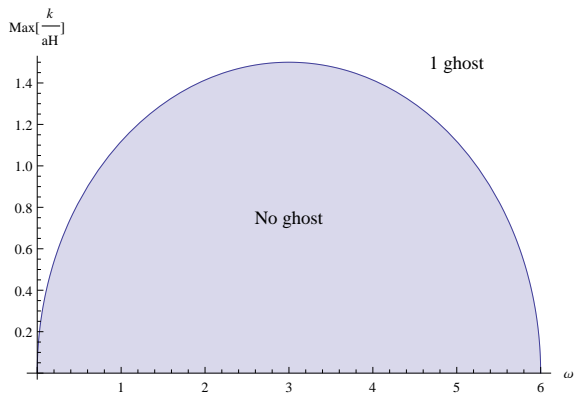}
\caption{The stability of the scalar sector implied by the determinant of the kinetic matrix~(\ref{detK}). For modes with $k/(aH)$ below the solid line, the determinant is positive, so there no ghost degrees of freedom~(see Eq.(\ref{qmg-newbasis}) for the field basis in which this is manifest). On the other hand, above the solid line, one degree of freedom has a positive kinetic term while the other is a ghost.}
\label{scareg}
\end{figure}

The scalar sector should consist of two dynamical degrees of freedom: the scalar field and the longitudinal mode of the massive graviton. 
The perturbations $\Phi$ and $B$ stemming from  $\delta g_{0i}$ and $\delta g_{00}$ are free of time derivatives, and so we eliminate them as auxiliary fields using their equations of motion:
\begin{equation}
B = \frac{r^2-1}{3\,\omega\,a\,H^2}\,
\left[
3\,H\,(\omega\,\delta\sigma-2\,\Phi)+\frac{1}{N}\left(k^2\,\dot{E} + 6\,\dot{\Psi}\right)\,
\right] \ ,
\end{equation}
\begin{eqnarray}
\Phi &=& \frac{1}{3\left[\omega\,(6-\omega)a^2\,H^2+4\,k^2\,(r^2-1)\right]}
\left[k^4\,\omega\,E + 3\,\omega\,\left(2\,k^2(r^2-1)-\frac{3\,\omega\,a^2\,H^2}{r-1}\right)\delta\sigma
\right.\nonumber\\
&&
\left.+3\,\omega\,\left(2\,k^2 +\frac{3\,\omega\,a^2\,H^2}{r-1}\right)\Psi- \frac{3\,\omega\,a^2\,H}{N}\left(\omega\,\delta\dot{\sigma}-6\,\dot{\Psi}\right) + \frac{2\,k^2}{H\,N}\,(r^2-1)\,(k^2\,\dot{E} +6\,\dot{\Psi})\right]\,.
\end{eqnarray}
Inserting these back into the action, we obtain an action with three fields, $\Psi$, $E$ and $\delta\sigma$. Since the ``sixth'' degree of freedom
(which would come from the Boulware-Deser instability in generic massive
theories) is removed by construction, there is another non-dynamical
combination, which we determine to be 
\begin{equation}
\tilde{\Psi} = \frac{1}{\sqrt{2}}\,\left(\Psi + \delta\sigma\right)\,.
\end{equation}
We also define an orthogonal combination,
\begin{equation}
\tilde{\delta\sigma} = \frac{1}{\sqrt{2}\,k^2}\,\left(\Psi-\delta\sigma\right)\,.
\end{equation}
With these field redefinitions, the action can be written in terms of $\tilde{\Psi}, \tilde{\delta\sigma}$ and $E$, with no time derivatives on $\tilde{\Psi}$. The latter is therefore auxiliary and can be eliminated via its equation:
\begin{eqnarray}
\tilde{\Psi} &=& \left(
-k^2 - \frac{24\,a^2\,H^2}{r\,(r-1)}+ \frac{2\,a^2\,H^2\,k^2\,\left\{\left[48-(6-\omega)\omega\right]r-\omega^2\right\}}{\left[4\,k^2-(6-\omega)\omega\,a^2\,H^2\right]\,(r-1)}
\right) \,\tilde{\delta\sigma}
-\frac{2\,\sqrt{2}\,k^4\,E}{3\,\left[4\,k^2-(6-\omega)\omega\,a^2\,H^2\right]}\nonumber\\
&&
+2\,a^2\,H\,\left(\frac{3}{r} + \frac{(6-\omega)\,[2\,k^2(r-1)+3\,\omega\,a^2\,H^2]}{\left[4\,k^2-(6-\omega)\omega\,a^2\,H^2\right](r-1)}\right)\frac{\dot{\tilde{\delta\sigma}}}{N}
+\frac{\sqrt{2}\,(6-\omega)\,k^2\,a^2\,H}{3\,\left[4\,k^2-(6-\omega)\omega\,a^2\,H^2\right]}\,\frac{\dot{E}}{N}\,.
\end{eqnarray}
Using this solution in the action, and introducing the notation $Y \equiv (\tilde{\delta\sigma},\; E)$, the scalar action can then be written as
\begin{equation}
S = \int \frac{d^3k}{2}\,a^3\,N\,dt\,\left[ \frac{\dot{Y}^\dagger}{N} \,{\cal K}\,\frac{\dot{Y}}{N} + \frac{\dot{Y}^\dagger}{N} \,{\cal M}\,Y + Y^\dagger\,{\cal M}^T\,\frac{\dot{Y}}{N}- Y^T\,\Omega^2\,Y\right]\,,
\label{eq:actionscaqmg}
\end{equation}
where ${\cal M}$ is a real anti-symmetric $2\times2$ matrix, and ${\cal K}$ and $\Omega^2$ are real symmetric $2\times2$ matrices. (Note that there is no loss of generality in taking ${\cal M}$ anti-symmetric, since the symmetric part can be absorbed into $\Omega^2$ by adding total derivatives). For now, we focus on the kinetic terms. The components of the matrix ${\cal K}$ are
\begin{eqnarray}
{\cal K}_{11} &=& 2\,k^4\,M_p^2\,\omega\,\left[1+\frac{9\,a^2\,H^2}{k^2\,(r-1)^2}- \frac{a^2\,H^2\,\left[\omega+(6-\omega)r\right]^2}{\left[4\,k^2-(6-\omega)\omega\,a^2\,H^2\right]\,(r-1)^2}\right]\,,\nonumber\\
{\cal K}_{12} &=& \sqrt{2}\,k^4\,M_p^2\,\omega\left[\frac{r}{\omega\,(r-1)}-
\frac{2\,k^2\,\left[\omega+(6-\omega)r\right]}{3\,\omega\,\left[4\,k^2-(6-\omega)\omega\,a^2\,H^2\right]\,(r-1)}\right]\,,\nonumber\\
{\cal K}_{22} &=& \frac{k^4\,M_p^2\,\omega}{36}\left[1- \frac{(6-\omega)^2\,a^2\,H^2}{4\,k^2-(6-\omega)\omega\,a^2\,H^2}\right]\, .
\label{eq:qmg-scalarkin}
\end{eqnarray}
For the case at hand, it is sufficient to study the determinant of the kinetic matrix ${\cal K}$ to determine the sign of the eigenvalues.\footnote{The absence of ghosts requires that both the determinant and the trace are positive. On the other hand, only one of these being negative is enough to deduce the existence of a ghost degree, which happens to be the case for the current system. For a detailed diagonalization treatment, we refer the reader to Appendix \ref{app:qmg-scalar}. } 
The determinant takes the comparatively simple form, 
\begin{equation}
{\det}\,{\cal K} = \frac{3\,M_p^4\,k^6\,\omega^2\,a^4\,H^4}{\left[\omega\,a^2\,H^2- \frac{4\,k^2}{6-\omega}\right]\,(r-1)^2}\,.
\label{detK}
\end{equation}
The sign of the determinant is determined by the sign of the quantity within the square brackets. First note that the determinant is
always negative if $\omega<0$. Along with the condition for a realistic cosmology obtained from (\ref{friedmann}), the range of allowed $\omega$ is thus 
\begin{equation}
0 < \omega <6\,,
\end{equation}
in agreement with \cite{D'Amico:2012zv}.  In order to have no ghosts in the scalar sector, we need (See Figure \ref{scareg})
\begin{equation}
\frac{k}{a\,H} < \frac{\sqrt{\omega(6-\omega)}}{2}\,.
\label{noghostsca}
\end{equation}

Generically, the right hand side of the inequality (\ref{noghostsca}) is
of order 1. This implies that for
modes with physical wavelengths that are smaller than cosmological
scales, one of the two degrees of freedom is a ghost. In other words, parametrically, there is an 
instability in the scalar sector at physical momenta above $H\sim m_g$. As shown in Appendix~\ref{app:qmg-scalar},
both the physical momenta and the energies of those ghost modes near
the threshold are not parametrically higher than $H\sim m_g$ and thus
are below the UV cutoff scale of the effective field theory. This
signals the presence of ghost instabilities in the regime of validity of the
effective field theory.

We end this section by comparing our results to those found in \cite{D'Amico:2012zv,Haghani:2013eya}.  Noting that at the level of the kinetic matrix (\ref{eq:qmg-scalarkin}) the only scale other than $H$ is the momentum, the limit $H\to0$ is equivalent to considering modes with wavelengths much shorter than the Hubble radius, i.e. $k \gg aH $, which is in contradiction with the no-ghost condition (\ref{noghostsca}). In this limit, the kinetic matrix then becomes
\begin{equation}
{\cal K} = M_p^2\,\omega\,k^4\,\left(
\begin{array}{cc}
2 & \frac{1}{3\,\sqrt{2}}\\
\frac{1}{3\,\sqrt{2}}& \frac{1}{36}
\end{array}
\right)
+ {\cal O}\left(\frac{a\,H}{k}\right)^{2}\,,
\end{equation}
which has one positive and one zero eigenvalue, as in \cite{D'Amico:2012zv} (See Appendix \ref{app:comparison} for a more detailed comparison). In other words, the apparent stability of the self-accelerating solution is due to the loss of the dynamics of the ghost degree of freedom in the short wave-length limit, and so the decoupling limit is not sufficient to determine stability, in agreement with~\cite{D'Amico:2012zv}.  In \cite{Haghani:2013eya}, only the super-horizon limit $k\rightarrow 0$ is considered, so the instability which appears only for physical wavelengths $\lesssim$ Hubble is not visible in this limit.

\subsection{Higher derivative terms and UV sensitivity}

The quasi-dilaton theory is governed by the global scaling symmetry
described in \cite{D'Amico:2012zv}.  The action \eqref{quasiaction}
includes all possible terms compatible with the symmetry, with up to two
derivatives, and we found there was no way to render the scalar
perturbations of the self-accelerating solutions stable at all momenta.  

However, beyond two derivative order there are many more terms
compatible with the symmetry.  These higher derivative terms can be
thought of as encoding UV effects from whatever physics completes the
theory.  Among the possible higher-derivative terms, we will focus here on two
classes of distinguished interaction terms which will not add new
degrees of freedom.  There are the
Goldstone-like terms of the form $\sim (\partial\sigma)^n$, and the
three possible non-trivial covariantized Galileon terms, of the form
$\sim(\partial\sigma)^2(\partial^2\sigma)^n+\cdots$
\cite{Deffayet:2009wt,Goon:2011qf,Goon:2011uw,Burrage:2011bt}. The
strong coupling scale of the quasi-dilaton on flat space is $\Lambda_3 \sim
( M_p m_g^2)^{1/3}$ \cite{D'Amico:2012zv}, so it is natural for the
Galileon-like terms to appear suppressed by this scale.   
The Goldstone-like terms should carry the scale\footnote{The
reason the Goldstone-like terms carry a higher scale is because the
$\Lambda_3$ decoupling limit of the theory has an enhanced Galilean
symmetry \cite{D'Amico:2012zv}, which the
Goldstone-type interactions
are not invariant under.  This means that they will not be generated in the
decoupling limit, so whatever the quantum corrections to these operators
are in the full theory, they should not survive in the decoupling limit,
i.e. they should be suppressed by a scale higher than $\Lambda_3$.}
$\Lambda_2 \sim  (M_p m_g)^{1/2}\gg \Lambda_3$. One can repeat the
calculation of the perturbations including these terms, in the hopes
that the fluctuations can be stabilized at short scales $k\gg H$.   

The Friedmann equation now becomes: 
\be
\left(3-\frac{\omega}{2}+3g_3h^2\right)H^2 
+ \frac{1}{2}m_g^2\left[f(h^2)-2h^2f'(h^2)\right] 
= \Lambda + \Lambda_{\rm SA}, \quad
h \equiv \frac{H}{m_g} \ ,
\ee
where $g_3$ is the dimensionless coupling for the cubic covariant 
Galileon and we have chosen the form $\Lambda_2^4\, f(x)/2$ as the 
Goldstone-like term, where $x=-(\nabla\sigma)^2/\Lambda_2^4$. 
On the other hand, the constraint equation and the value of
$\Lambda_{\rm SA}$ remain the same.  (We have omitted the quartic and
quintic Galileon terms for simplicity.) There are still self-accelerating
solutions with $H\sim m_g$ so the existence of these solutions appears 
insensitive to the UV effects encoded by the higher derivative
operators. For a sensible cosmology, $H^2$ determined by the Friedmann
equation should be an increasing function of the bare cosmological
constant $\Lambda$ (which represents the matter energy density in our setup). Demanding this, we obtain the condition
\be
6 - \omega + 12g_3h^2 - f'(h^2)-2h^2f''(h^2) > 0\ . 
\label{eqn:GNcosmopositivity}
\ee
The determinant of the kinetic matrix ${\cal K}$ for scalar fluctuations
changes by order one, 
\be
\det{\cal K} = \frac{3M_p^4k^6[\omega-3g_3h^2+f'(h^2)]^2a^4H^4}
{\left\{[\omega-3g_3h^2+f'(h^2)]a^2H^2
-\frac{(2+g_3h^2)^2k^2}{6-\omega+12g_3h^2-f'(h^2)-2h^2f''(h^2)}
\right\}(r-1)^2}\ ,
\ee
but the determinant is still always negative for sufficiently large
momenta, provided that the condition (\ref{eqn:GNcosmopositivity}) is
satisfied.

The quartic and quintic covariant Galileon terms can render the determinant of the
kinetic matrix for scalar perturbation positive for large momenta,
depending on the values of the coupling constants. However, in this regime
of parameters, the tensor and vector modes acquire negative kinetic terms. Moreover, 
after explicit diagonalization of the kinetic matrix one
can show that there are two ghost modes in the scalar sector, provided that $H^2$, determined
by the Friedmann equation, is an increasing function of
$\Lambda$. 

Thus, the form of the dispersion relations for the perturbations depends
on and receives order one correction due to UV effects, but the presence
of the ghost seems to be a robust feature.

\section{Varying mass theory}

We now turn to the varying mass theory, obtained by introducing a scalar into dRGT theory and allowing the graviton mass to be a function of this scalar.  This theory was first considered in \cite{D'Amico:2011jj}, and further studied in \cite{Huang:2012pe}.

The action is
\begin{equation}
S= \int\,d^4x\,\sqrt{-g}\,\left\{{M_p^2\over 2}\Big[{R}[g]- 2\Lambda+ 2m_g^2(\sigma) \left[{\cal L}_2 +\alpha_3(\sigma)\,{\cal L}_3 +\alpha_4(\sigma)\,{\cal L}_4\right]\Big]-\frac{1}{2}\,\partial_\mu\sigma\,\partial^\nu\sigma-V(\sigma)\right\}\,.
\end{equation}
We have further generalized to allow the dRGT parameters $\alpha_3$ and $\alpha_4$ to depend on the scalar $\sigma$.  
The part of the action which provides mass to the graviton takes the same form as in Eq.(\ref{lag234}), but here the building block tensor ${\cal K}$ is the same as in dRGT theory~\cite{deRham:2010kj}, i.e.
\begin{equation}
 {\cal K}^\mu _\nu = \delta^\mu _\nu 
 - \left(\sqrt{g^{-1}\bar g}\right)^{\mu}_{\ \nu}\,.
\label{Kdefmasvar}
\end{equation}

One of our goals is to compare our results with the analogous analysis of perturbations in the dRGT theory. Since the original theory does not allow flat solutions for a Minkowski reference metric, it is necessary to adopt a more general form. We therefore extend the fiducial metric to be an arbitrary spatially flat homogeneous and isotropic metric,
\begin{equation}
\bar g_{\mu\nu}dx^{\mu}dx^{\nu} = -n(t)^2dt^2+ \alpha(t)^2 \delta_{ij}dx^idx^j.
\end{equation}

\subsection{Cosmological Background Equations }

We first study the cosmological background equations (see \cite{Huang:2012pe,Saridakis:2012jy,Cai:2012ag,Hinterbichler:2013dv,Wu:2013ii,Leon:2013qh} for more on background cosmological solutions to mass-varying massive gravity.)
For the physical background metric, we adopt the flat FRW ansatz
\begin{equation}
g_{\mu\nu} dx^\mu\,dx^\nu = -N(t)^2\,dt^2+a(t)^2\,\delta_{ij} dx^i\,dx^j\,.
\label{flatfrw2}
\end{equation}

To write the mini-superspace action, we introduce time reparametrization via a St\"uckelberg field $f(t)$, by replacing the fiducial metric with
\begin{equation}
\bar g_{\mu\nu}dx^{\mu}dx^{\nu} = -n(f(t))^2f'(t)^2dt^2+ \alpha(f(t))^2 \delta_{ij}dx^idx^j.
\end{equation}
Unitary gauge corresponds to the choice $f(t)=t$.  

The mini-superspace action is
\begin{align}
	\frac{S}{V}
		& = \int{\rm d}t\, \left\{3 M_{\rm P}^2 \left[ - \frac{\dot{a}^2a}{N}+m_g^2\left(NF - \dot{f} n(f)G\right)\right]+a^3\left[ \frac{1}{2} N^{-1}\dot{\sigma}^2 - NV(\sigma) - N\,M_{\rm P}^2\,\Lambda \right]\right\},
\end{align}
where $V$ is the comoving volume and 
\begin{align} 
	F
		& \equiv a(a-\alpha(f))(2a-\alpha(f)) + \frac{\alpha_3}{3}(a-\alpha(f))^2(4a-\alpha(f)) + \frac{\alpha_4}{3}(a-\alpha(f))^3, \label{Fofa}
 \\
	G
		& \equiv a^2(a-\alpha(f))+\alpha_3 a(a-\alpha(f))^2 + \frac{\alpha_4}{3}(a-\alpha(f))^3.\label{Gofa}
\end{align}

In the following, for clarity, we will use the definitions
\begin{equation}
H\equiv {\dot a\over N a}\, , \qquad X \equiv \frac{\alpha}{a}\,,
\qquad
\bar H \equiv \frac{\dot{\alpha}}{n\,\alpha}\,,
\qquad
r\equiv \frac{n}{N\,X}\,,
\end{equation}
and we will omit the dependence of the functions $m_g$, $\alpha_3$, $\alpha_4$ and $V$ on the field value $\sigma$. (We also caution the reader that the above definitions of $X$ and $r$ are different than the ones in the quasi-dilaton theory, which we introduced in Section \ref{quasdisec}.)

The equation of motion for the temporal St\"uckelberg field $f$ is
\begin{eqnarray}
-{1\over 3M_p^2Nn}\left.\frac{\delta\,S}{\delta\,f}\right|_{f=t} =&&\frac{1}{N}\frac{d}{dt}
\Big\{m_g^2
a^3\,(X-1)\,
\left[
1-\,(X-1)\,\alpha_3+{1\over 3}(X-1)^2\,\alpha_4\right]
\Big\}
\nonumber\\
&&\qquad
+\,a^3\,\bar H\,m_g^2\,X\,
\Big[
3-X(2+r)+(X-1) ((1+2r)X-3)\alpha_3-( X-1)^2 ( r X-1)\alpha_4  \Big]=0. \nn\\
\,
\label{eqstuck-mv}
\end{eqnarray}

The Friedmann equation is obtained by varying the action with respect to $N$, 
\begin{equation}
{1\over M_p^2a^3}\left.\frac{\delta\,S}{\delta\,N}\right|_{f=t} =3\,H^2 -\Lambda -\frac{1}{M_p^2}\,\left(\frac{\dot{\sigma}^2}{2\,N^2}+V\right) -m_g^2\,(X-1)\,\left[-3 (X-2 ) + (X-4 ) (X-1 ) \alpha_3 + (X-1 )^2 \alpha_4\right]=0\, ,
\label{eqn-vm}
\end{equation}
and by taking a variation with respect to $a$, we obtain the dynamical equation which, after forming a linear combination with (\ref{eqn-vm}), can be expressed as
\begin{equation}
{1\over 3M_p^2Na^2}\left.\frac{\delta\,S}{\delta\,a}\right|_{f=t} -{1\over M_p^2a^3}\left.\frac{\delta\,S}{\delta\,N}\right|_{f=t} =\frac{2\,\dot{H}}{N}
+\frac{\dot{\sigma}^2}{M_p^2\,N^2}-
m_g^2\,(r-1)\,X\left[3 - 2 X + (X-3 ) (X-1) \alpha_3 + (X-1)^2 \alpha_4\right]=0
\, .
\label{eqa-vm}
\end{equation}

Finally, the equation of motion for $\sigma$ is
\begin{eqnarray}
-{1\over a^3N}\left.\frac{\delta\,S}{\delta\,\sigma}\right|_{f=t} =&&\frac{1}{N}\,\frac{d}{dt}\,\left(\frac{\dot{\sigma}}{N}\right)+3\,H\,\frac{\dot{\sigma}}{N}+V'
\nonumber\\
&&\qquad-M_p^2\,m_g^2\,(X-1)^2\Bigg\{
\alpha_3'\,(4-X(1+3\,r))+\alpha_4'(X-1)\,(r\,X-1)
\nonumber\\
&&\qquad\qquad+\frac{2\,m_g'}{m_g}
\Bigg[\frac{3 (X (r+1)-2)}{X-1} - ( X(1+ 3 r )-4) \alpha_3 + (X-1 ) ( r X-1) \alpha_4\Bigg]\Bigg\}=0\,,
\label{eqs-vm}
\end{eqnarray}
where a prime denotes differentiation with respect to $\sigma$.

It is convenient to cast these equations into a more familiar perfect fluid-like form, by defining the following quantities
\begin{eqnarray}
\rho_m &\equiv& M_p^2 m_g^2\,(X-1)\,\left[-3 (X-2)+ (X-4) (X-1)\alpha_3+ (X-1)^2\alpha_4\right]\,,
\nonumber\\
p_m &\equiv& M_p^2 m_g^2\,\left[6 - 3X( r+2 ) + X^2 (1+2r) - (X-1 ) (4 -  X (3r+2) + r X^2) \alpha_3 - (X-1  
    )^2 (rX-1 ) \alpha_4\right]\,,
\nonumber\\
Q&\equiv& 
M_p^2\,m_g^2\,\frac{\dot{\sigma}}{N}\,(X-1)^2\Bigg\{
\alpha_3'\,(4-X(1+3\,r))+\alpha_4'(X-1)\,(r\,X-1)
\nonumber\\
&&\qquad\qquad\qquad\qquad\qquad\qquad+\frac{2\,m_g'}{m_g}\,
\Bigg[\frac{3 (X (r+1)-2)}{X-1} - ( X(1+ 3 r )-4) \alpha_3 + (X-1 ) ( r X-1) \alpha_4\Bigg]\Bigg\}\,,
\nonumber\\
\rho_\sigma &\equiv&  \frac{\dot{\sigma}^2}{2\,N^2}+V \,, \nn\\
 p_\sigma &\equiv&  \frac{\dot{\sigma}^2}{2\,N^2}-V \,,
\end{eqnarray}
in terms of which Eqs.~(\ref{eqstuck-mv})-(\ref{eqs-vm}) can be re-written, respectively, as
\begin{eqnarray}
&& \frac{\dot{\rho}_m}{N}+3\,H\,(\rho_m+p_m)=-Q\,,\nonumber\\
&&3\,H^2 = \Lambda+ \frac{1}{M_p^2}\,\left(\rho_\sigma + \rho_m \right)\,, \nonumber\\
&&\frac{\dot{H}}{N} = -\frac{1}{2\,M_p^2}\,\left[ (\rho_\sigma+p_\sigma) +(\rho_m+p_m)\right]\,, \nonumber\\
&& \frac{\dot{\rho}_\sigma}{N}+3\,H\,(\rho_\sigma+p_\sigma)=Q\,.
\end{eqnarray}
From here on, we will use these forms of the cosmological background equations.

\subsection{Tensor perturbations}
We proceed with the perturbation theory in the same manner as described in Section \ref{pertsub}.  We start with tensor perturbations around the background (\ref{flatfrw2}),
\begin{equation}
\delta g_{ij}=a^2h_{ij}^{TT}\,,
\end{equation}
where $\partial^ih_{ij}^{TT} = h^{TT\;i}_i=0$. 

The action for the tensor perturbations reads 
\begin{equation}
S= \frac{M_p^2}{8}\,\int d^3k\,a^3\,N\,dt\,\left(\frac{1}{N^2}|\dot{h}^{TT}_{ij}|^2-\left( \frac{k^2}{a^2} + M_{GW}^2\right)|h^{TT}_{ij}|^2\right)\,,
\end{equation}
where the mass term is
\begin{equation}
M_{GW}^2=\frac{(r-1)\,X^2}{(X-1)^2}\,\left[m_g^2\,(X-1) - \frac{\rho_m}{M_p^2}\right]
-\left(\frac{1}{r-1}+\frac{2\,X}{X-1}\right)
\,\frac{\rho_m+p_m}{M_p^2}\,.
\label{mgwdef}
\end{equation}
To obtain this, we have used the background acceleration equation \eqref{eqa-vm}.
In the case of self accelerating solutions \cite{Gumrukcuoglu:2011ew,Gumrukcuoglu:2011zh} of the standard dRGT theory, i.e. when $\dot{\sigma} = 0$ and $\rho_m = - p_m= {\rm const.}$, the last term in \eqref{mgwdef} drops out of the calculation, and the mass reduces to the one found in \cite{Gumrukcuoglu:2011zh}. We also stress that $M_{GW}$ here is different than the one defined for the quasi-dilaton theory in Section \ref{quasdisec}.

\subsection{Vector perturbations} 
\label{sec:vecvar}
Next we consider transverse vector perturbations to the metric
\begin{equation}
\delta g_{0i} = N\,a\,B^T_i\,,
\qquad
\delta g_{ij} = \frac{a^2}{2}\,(\partial_i E^T_j+\partial_j E^T_i)\,,
\end{equation}
where $\partial^iB^T_i=\partial^iE^T_i=0$.  The field $B_i$ appears without time derivatives and may be eliminated as an auxiliary field using its own equation of motion,
\begin{equation}
B^T_i = \frac{a}{2\,\left[1-\frac{2\,a^2}{k^2\,M_p^2\,(r^2-1)}\,(\rho_m+p_m)\right]}\,\frac{\dot{E}^T_i}{N}\,.
\end{equation}

Once this solution is inserted back into the action, what remains is an action for one dynamical vector
\begin{equation}
S= \frac{M_p^2}{8}\,\int d^3k\,a^3\,N\,dt\,\left(\frac{{\cal T}_V }{N^2}|\dot{E}^T_{i}|^2-\frac{k^2\,M_{GW}^2}{2}|E^T_{i}|^2\right)\,,
\end{equation}
where
\begin{equation}
{\cal T}_V \equiv \frac{k^2}{2} \,\left(1-\frac{k^2\,(r^2-1)\,M_p^2}{2a^2\,(\rho_m+p_m)}\right)^{-1}\,,
\label{kinvec-vm}
\end{equation}
and the tensor mode mass $M_{GW}$ is as defined as in (\ref{mgwdef}).  For the self-accelerating solutions of the dRGT theory, where $\dot{\sigma}=0$ and $\rho_m+p_m= 0$, the vector kinetic term vanishes, in agreement with the results of \cite{Gumrukcuoglu:2011zh}.

From Eq.(\ref{kinvec-vm}), we see that for $(r^2-1)/(\rho_m+p_m)>0$
there is a critical momentum scale above which the vector modes become
ghosts. On the other hand, in the opposite case with
$(r^2-1)/(\rho_m+p_m)\leq 0$, the timelike kinetic term of vector modes
always has the correct sign, and thus there is no such critical momentum
scale. The stability of the system thus requires that such a critical
momentum scale should be either absent or above $\Lambda_{UV}$, the UV cutoff scale of
the effective field theory, i.e. 
\begin{equation}
\frac{\Lambda_{UV}^2\,(1-r^2)}{H^2R} < 2,\quad
 R \equiv -\frac{\rho_m+p_m}{M_p^2H^2} \ .
\label{noghostvec-vm}
\end{equation}

In order to determine the stability conditions and the time scale of potential instabilities, it is useful to use canonical normalization. However, the existence of several unknown functions and the lack of a simple background prevent us from performing the stability analysis in a complete way. On the other hand, assuming that the tensor modes have positive squared-mass (\ref{mgwdef}), i.e. $M_{GW}^2>0$ and the vector sector is free of ghosts (\ref{noghostvec-vm}), we can still obtain sufficient (but not necessary) conditions to ensure the stability of the modes.  
These conditions cause the vector modes to damp with time, so tachyon-like instabilities can be avoided. It is convenient to perform a time reparametrization choosing the lapse function to be 
$N=a^3{\cal T}_V$ 
\begin{equation}
S|_{N=a^3{\cal T}_V}= \frac{M_p^2}{8}\,\int d^3k\,dt\,
 \left(|\dot{E}^{T}_{i}|^2-\frac{k^2}{2}a^6{\cal T}_VM_{GW}^2|E^{T}_{i}|^2\right)\,.
\end{equation}
Hence, provided that the conditions $M_{GW}^2>0$ and ${\cal T}_V>0$ are already imposed,
the amplitudes of the variables $E^{T}_i$ decrease as the
universe expands if 
\begin{equation}
 \partial_t\left[a^6{\cal T}_VM_{GW}^2\right] > 0 \ .
\end{equation}
Demanding that this condition holds for all momenta below $\Lambda_{UV}$, we
obtain 
\begin{equation}
 {\cal A}\,\frac{\Lambda_{UV}^2(1-r^2)}{H^2R} < \frac{3}{2}{\cal B}\,,
\label{stabilityvm}
\end{equation}
where
\begin{equation}
{\cal A} = 1 + \frac{1}{8NH}\frac{d}{dt}
 \ln\left(\frac{RM_{GW}^2}{r^2-1}\right), \quad
 {\cal B} = 1 + \frac{1}{6NH}\frac{d}{dt}\ln \left(M_{GW}^2\right) \ .
\end{equation}

\subsection{Scalar perturbations}
\label{vm-scalar}
We now move on to the scalar perturbations. In the absence of matter, we expect the sector to contain two degrees of freedom. The scalar parts of the metric perturbations are
\begin{equation}
\delta g_{00} = -2\,N^2\,\Phi\,,
\qquad
\delta g_{0i} = N\,a\,\partial_i B\,,
\qquad
\delta g_{ij} = a^2\,\left[
2\,\delta_{ij}\,\Psi +\left(\partial_i\partial_j - {1\over 3}{\delta_{ij}}\,\partial_l\partial^l\right)E\right]\,,
\end{equation}
while the scalar field is expanded as,
\begin{equation}
\sigma = \sigma^{(0)} +M_p\,\delta\sigma \ .
\end{equation}
The perturbations $\Phi$ and $B$ coming from $\delta g_{0i}$ and $\delta g_{00}$ carry no time derivatives and are non-dynamical, so we may determine them using their own equations of motion:
\begin{eqnarray}
B&=& - \frac{M_p^2\,(r+1)}{a\,\left[4\,M_p^4\,H^2\,\tfrac{k^2}{a^2}\,(r^2-1)+(\rho_m+p_m)(\rho_\sigma+p_\sigma-6\,M_p^2\,H^2)\right]}
\nonumber\\
&&\times\Bigg\{
\frac{k^2\,(r-1)}{3}\, \left[2\,M_p^2\,H\,\frac{k^2}{a^2}\,E +(\rho_\sigma+p_\sigma-6\,M_p^2\,H^2)\,\frac{\dot{E}}{N}\right]
\nonumber\\
&&\quad+2\,H\,\left[2\,M_p^2\,\tfrac{k^2}{a^2}\,(r-1)-3\,(\rho_m+p_m)\right]\Psi + 2\,(r-1)(\rho_\sigma +p_\sigma)\,\frac{\dot{\Psi}}{N}-\frac{2\,M_p\,H\,(r-1)\,(\rho_\sigma+p_\sigma)}{\dot{\sigma}}\,\delta\dot{\sigma}
\nonumber\\
&&\quad+\left[\frac{M_p\,H\,N}{\dot{\sigma}}\,\left( 6\,(H-\bar H\,r\,X)(\rho_m+p_m)- \frac{r-1}{N}\,(\dot{\rho}_\sigma-\dot{p}_\sigma+2\,\dot{\rho}_m)\right)+\frac{(r-1)(\rho_\sigma+p_\sigma-6\,M_p^2\,H^2)\,\dot{\sigma}}{M_p\,N}\right]\,\delta\sigma\Bigg\}\,,
\nonumber\\
\Phi&=& - \frac{M_p^2}{4\,M_p^4\,H^2\,\tfrac{k^2}{a^2}\,(r^2-1)+(\rho_m+p_m)(\rho_\sigma+p_\sigma-6\,M_p^2\,H^2)}
\nonumber\\
&&\times\Bigg\{
\frac{k^4}{3\,a^2}\, \left[(\rho_m+p_m)\,E -2\,M_p^2\,H(r^2-1)\,\frac{\dot{E}}{N}\right]
+\frac{\rho_m+p_m}{M_p^2\,(r-1)}\,
\left[2\,M_p^2\,\tfrac{k^2}{a^2}\,(r-1)-3\,(\rho_m+p_m)\right]\Psi 
\nonumber\\
&&\quad
-2\,H\,
\left[2\,M_p^2\,\tfrac{k^2}{a^2}\,(r^2-1)-3\,(\rho_m+p_m)\right]
\frac{\dot{\Psi}}{N}-\frac{(\rho_m+p_m)(\rho_\sigma + p_\sigma)}{M_p\,\dot{\sigma}}\,\delta\dot{\sigma}
\nonumber\\
&&\quad+\left[\frac{(\rho_m+p_m)\,N}{2\,M_p^2\,(r-1)\dot{\sigma}}\,\left( 6\,(H-\bar H\,r\,X)(\rho_m+p_m)- \frac{r-1}{N}\,(\dot{\rho}_\sigma-\dot{p}_\sigma+2\,\dot{\rho}_m)\right)-\frac{2\,M_p\,H\,k^2\,(r^2-1)\,\dot{\sigma}}{a^2\,N} \right]\,\delta\sigma
\Bigg\}\,.
\end{eqnarray}
Inserting these back into the action, we end up with a system of three
degrees of freedom, $\Psi$, $E$ and $\delta\sigma$. Since the would-be Boulware-Deser ghost is removed by construction, there is another non-dynamical combination, which is found to be 
\begin{equation}
\tilde{\Psi} = \frac{1}{\sqrt{2}}\,\left(\Psi + \frac{M_p\,H\,N}{\dot{\sigma}}\delta\sigma\right)\,.
\end{equation}
We also define an orthogonal combination,
\begin{equation}
\tilde{\delta\sigma} = \frac{1}{\sqrt{2}\,k^2}\,\left(\Psi - \frac{M_p\,H\,N}{\dot{\sigma}}\delta\sigma\right)\,.
\end{equation}
The action can now be written in terms of
$\tilde{\Psi}, \tilde{\delta\sigma}$ and $E$, with no time derivatives on
$\tilde{\Psi}$. The latter is auxiliary and can be eliminated with its own equation of motion.  Thus, we
obtain an action in terms of $\tilde{\delta\sigma}$ and $E$ of the form 
\begin{equation}
S = \int \frac{d^3k}{2}\,a^3\,N\,dt\,\left[ \frac{\dot{Y}^\dagger}{N} \,{\cal K}\,\frac{\dot{Y}}{N} + \frac{\dot{Y}^\dagger}{N} \,{\cal M}\,Y + Y^\dagger\,{\cal M}^T\,\frac{\dot{Y}}{N}- Y^\dagger\,\Omega^2\,Y\right]\,,
\end{equation}
where $Y \equiv (\tilde{\delta\sigma},\; E)$, ${\cal M}$ is a real
$2\times2$ matrix, and ${\cal K}$ and $\Omega^2$ are real symmetric
$2\times2$ matrices. Note that by adding boundary terms, the mixing
matrix ${\cal M}$ between fields and derivatives can be made
anti-symmetric. 

The full kinetic matrix is rather lengthy, and so we will not display the full expression, other than to note that for the dRGT theory, where $\dot{\sigma}=0$ and $\rho_m = - p_m$, it can be checked that ${\cal K}=0$, consistent with the results of \cite{Gumrukcuoglu:2011zh}.  Specializing to a Minkowski reference metric ($\bar H = \dot{\bar H}= 0$) brings the kinetic matrix to a more manageable form:
\begin{eqnarray}
{\cal K}_{11} &=& \frac{a^2k^2}{(r-1)^2}\,\Bigg\{ 18\,(\rho_m+p_m) \,\left(\frac{2\,M_p^2k^2r^2}{2\,M_p^2k^2+3\,a^2\,(\rho_m+p_m)}-1\right)
\nonumber\\
&&\qquad\qquad\quad
+\frac{4\,M_p^2k^2\left[2\,M_p^2\,\tfrac{k^2}{a^2}\,(r-1)-3\,(\rho_m+p_m)\right]^2(\rho_\sigma+p_\sigma)}{[2\,M_p^2\,k^2+3\,a^2(\rho_m+p_m)]\,\left[4\,M_p^4\,H^2\,\tfrac{k^2}{a^2}-(\rho_m+p_m)(\rho_\sigma + p_\sigma -6\,M_p^2\,H^2)\right]}
\Bigg\}\,,
\nonumber\\
{\cal K}_{12} &=& 
\frac{\sqrt{2}\,M_p^2\,k^4}{3\,(r-1)}\,
\Bigg\{
\frac{9\,r\,(\rho_m+p_m)}{2\,M_p^2\,\tfrac{k^2}{a^2}+3\,(\rho_m+p_m)}
\nonumber\\
&&\qquad\qquad\quad
+\frac{2\,M_p^2k^2\left[2\,M_p^2\,\tfrac{k^2}{a^2}\,(r-1)-3\,(\rho_m+p_m)\right]\,(\rho_\sigma+p_\sigma)}{[2\,M_p^2\,k^2+3\,a^2(\rho_m+p_m)]\,\left[4\,M_p^4\,H^2\,\tfrac{k^2}{a^2}-(\rho_m+p_m)(\rho_\sigma + p_\sigma -6\,M_p^2\,H^2)\right]}
\Bigg\}\,,
\nonumber\\
{\cal K}_{22} &=& 
\frac{M_p^2\,k^4}{18}\,\Bigg\{
\frac{9\,(\rho_m+p_m)}{2\,M_p^2\,\tfrac{k^2}{a^2}+3\,(\rho_m+p_m)}
\nonumber\\
&&\qquad\qquad\quad
+\frac{4\,M_p^4k^4 (\rho_\sigma+p_\sigma)}{a^2[2\,M_p^2\,k^2+3\,a^2(\rho_m+p_m)]\,\left[4\,M_p^4\,H^2\,\tfrac{k^2}{a^2}-(\rho_m+p_m)(\rho_\sigma + p_\sigma -6\,M_p^2\,H^2)\right]}
\Bigg\}\,,
\label{eq:kinmatvm}
\end{eqnarray}
with determinant
\begin{equation}
{\rm det}[{\cal K}] =
\frac{3\,M_p^2\,a^2\,k^6\,(\rho_m+p_m)^2\,(\rho_\sigma+p_\sigma-6\,M_p^2\,H^2)}{(r-1)^2\,\left[4\,M_p^4\,H^2\,\tfrac{k^2}{a^2}-(\rho_m+p_m)(\rho_\sigma + p_\sigma -6\,M_p^2\,H^2)\right]}\,.
\label{eq:vm-noghostsca}
\end{equation}
We stress that we have not specified any background solution up to this point; the
only choice we have made is to fix the fiducial metric to be Minkowski. By requiring that the determinant is positive, we infer that in order to avoid a ghost degree of freedom, the momentum
should satisfy\footnote{Note that absence of ghosts requires that both the determinant and the trace are positive. However, for the scenario at hand, applying the field redefinitions given in Appendix \ref{app:diagonalkin} shows that one of the degrees of freedom always has a positive kinetic term. Thus, Eq.(\ref{eq:vm-noghostsca}) is enough to ensure a healthy kinetic action.}
\begin{equation}
\left(\frac{\rho_\sigma+p_\sigma}{4\,M_p^2\,H^2}-\frac{3}{2}\right)^{-1}\frac{k^2}{a^2} > \frac{\rho_m+p_m}{M_p^2}\,.
\label{eq:vm-noghostsca2}
\end{equation}
Note that this condition should be imposed for all $k$ in the regime
$0\leq k/a\leq \Lambda_{UV}$, where $\Lambda_{UV}$ is the UV cutoff
scale of the theory.

In a regime in which we have a de Sitter like expansion, i.e. 
$|{\dot H}| \ll H^2$, this condition becomes even simpler, 
\begin{equation}
 R + \frac{4}{R-6}\frac{k^2}{H^2a^2} > 0\ , 
\label{noghostsca-vm-k}
\end{equation}
where $R$ is defined in (\ref{noghostvec-vm}). Demanding that the condition 
(\ref{noghostsca-vm-k}) holds for all physical momenta $k/a < \Lambda_{UV}$ and supposing that $\Lambda_{UV}/H>3/2$, we
obtain the no-ghost condition for scalar perturbations in the regime
$|{\dot H}| \ll H^2$ as 
\begin{equation}
 R > 6 \ . 
\label{noghostsca-vm}
\end{equation}

\subsection{Consistency of stability conditions}

We now discuss the regions of parameter space in which the stability requirements we
obtained in Eqs.(\ref{mgwdef}), (\ref{noghostvec-vm}),
(\ref{stabilityvm}) and (\ref{noghostsca-vm}) can be satisfied. The
summary of the conditions is: 
\begin{enumerate}[{\it i.{\rm )}}]
 \item To avoid a tachyonic instability in the tensor sector, we need (from Eq.(\ref{mgwdef})) 
\begin{equation}
M_{GW}^2 > 0\,.
\end{equation}
\item To avoid a ghost instability in the vector sector, from Eq.(\ref{noghostvec-vm}),
\begin{equation}
\frac{\Lambda_{UV}^2\,(1-r^2)}{H^2R} < 2\,.
\end{equation}
\item To avoid the unchecked growth of vector perturbations, from Eq.(\ref{stabilityvm}),
\begin{equation}
 {\cal A}\,\frac{\Lambda_{UV}^2(1-r^2)}{H^2R} < \frac{3}{2}{\cal B}\,.
\end{equation}
\item To avoid a ghost instability in the scalar sector, from Eq.(\ref{noghostsca-vm}),
\begin{equation}
 R > 6\,.
\end{equation}
\end{enumerate}
Here, we have defined 
\begin{equation}
R \equiv -\frac{(\rho_m+p_m)}{H^2\,M_p^2}, \quad
{\cal A} = 1 + \frac{1}{8NH}\frac{d}{dt}
 \ln\left(\frac{RM_{GW}^2}{r^2-1}\right), \quad
 {\cal B} = 1 + \frac{1}{6NH}\frac{d}{dt}\ln \left(M_{GW}^2\right)\,,
\end{equation}
and have assumed that the UV cutoff scale $\Lambda_{UV}$ is higher than
$3/2$ in units of $H$ and that the expansion is de Sitter-like, i.e. 
$|\dot{H}| \ll H^2$ (relevant for the scalar sector no-ghost condition). 

Note that if we satisfy the condition {\it iv.}, then the condition
{\it ii.} is trivially satisfied if $r^2>1$, and that the condition {\it iii.} is
also trivially satisfied in this case if both ${\cal A}$ and ${\cal B}$
are positive. In more general cases, the above set of stability
conditions is less trivial, but in principle there are regimes in which all of them
are simultaneously satisfied.

\section{Discussion}
If the cosmologies of any of the recently proposed variations of massive gravity are to be of phenomenological use, it is crucial to understand the extent to which the theories propagate well-behaved, ghost free perturbations around their cosmological backgrounds. In this paper we have carried out this calculation for the cases of the quasi-dilaton theory and for the mass varying massive gravity theory. We find a host of constraints on these theories, primarily stemming from the requirement that ghost degrees of freedom not appear in the regime of applicability of the effective field theory.  In the case of the quasi-dilaton theory, it can be seen that the stability found in the decoupling limit is an artifact of that particular limit and that, in fact, a ghost degree of freedom remains in the full theory.

\begin{acknowledgments}
We thank the authors of \cite{D'Amico:2012zv} for valuable correspondence and comments.
The work of A.E.G, C.L. and S.M. was supported by the World Premier International Research Center Initiative (WPI Initiative), MEXT, Japan. S.M. also acknowledges the support by Grant-in-Aid for Scientific Research 24540256 and 21111006. The work of MT is supported in part by the US Department of Energy and NASA ATP grant NNX11AI95G.  Research at Perimeter Institute is supported by the Government of Canada through Industry Canada and by the Province of Ontario through the Ministry of Economic Development and Innovation. This work was made possible in part through the support of a grant from the John Templeton Foundation. The opinions expressed in this publication are those of the authors and do not necessarily reflect the views of the John Templeton Foundation (KH).  KH and MT thank the Institute for the Physics and Mathematics of the Universe (IPMU) at the University of Tokyo, where this collaboration began, for their wonderful hospitality.
\end{acknowledgments}

\appendix

\section{Dispersion relation for scalar perturbations in the quasi-dilaton theory}
\label{app:qmg-scalar} 

In this Appendix, we provide the details of the diagonalization procedure for the quadratic action for scalar perturbations in the quasi-dilaton theory, starting from (\ref{eq:actionscaqmg}).

\subsection{Canonical normalization}
We first introduce a new field basis
\begin{eqnarray}
Z_1 &\equiv& \frac{k^2\,M_p}{3\,\sqrt{\omega}}\,\left\vert \frac{4}{\omega\,(6-\omega)}-\frac{a^2\,H^2}{k^2}\right\vert^{-1/2}
\left[
E + 6\,\sqrt{2}\,\left(1+\frac{3\,\omega\,a^2\,H^2}{2\,(r-1)\,k^2}\right)\tilde{\delta\sigma}\right]\,,\nonumber\\
Z_2 &\equiv& \frac{k^2\,M_p}{\sqrt{6}}\,\left(E+6\,\sqrt{2}\,\tilde{\delta\sigma}\right)\, ,
\label{qmg-newbasis}
\end{eqnarray}
in terms of which the kinetic matrix (\ref{eq:qmg-scalarkin}) becomes diagonal and canonically normalized, so that the action is formally
\begin{equation}
S = \int \frac{d^3k}{2}\,a^3\,N\,dt\,\left[ \frac{\dot{Z}^\dagger}{N} \,{\cal K}\,\frac{\dot{Z}}{N} + \frac{\dot{Z}^\dagger}{N} \,{\cal M}\,Z - Z^\dagger\,{\cal M}\,\frac{\dot{Z}}{N}- Z^\dagger\,\Omega^2\,Z\right]\, ,
\end{equation}
with a canonical form for the kinetic matrix:
\begin{equation}
{\cal K} = \left(
\begin{array}{cc}
{\rm Sign}(1-\tilde{k}^2) & 0\\
0 & 1
\end{array}
\right)\,,
\end{equation}
where we have introduced the dimensionless (and time dependent) momentum via
\begin{equation}
\tilde{k} \equiv \frac{2\,k}{a\,H\,\sqrt{\omega\,(6-\omega)}}\,.
\end{equation}
By adding appropriate total derivatives, the mixing matrix ${\cal M}$ can be made antisymmetric, reading
\begin{equation}
{\cal M} = -\sqrt{1-\frac{\omega}{6}}\,H\,(2\,r-1)\frac{\tilde{k}\,\sqrt{\vert 1-\tilde{k}^2\vert}}{1-\tilde{k}^2}\,\left(
\begin{array}{cc}
0 & 1\\
-1 & 0
\end{array}
\right)\,.
\label{qmg-mixm}
\end{equation}
Finally, the components of the symmetric  matrix $\Omega^2$ are
\begin{eqnarray}
(\Omega^2)_{11} &=& -\frac{H^2}{4}\,\frac{\tilde{k}^2 \, \vert 1-\tilde{k}^2\vert}{1-\tilde{k}^2}
\left[(8\,r-\omega-2)(8\,r-\omega-8)+\frac{4\,(8\,r-\omega-6)}{1-\tilde{k}^2}+\frac{12}{(1-\tilde{k}^2)^2}\right]\,,\nonumber\\
(\Omega^2)_{12} &=&
-\sqrt{1-\frac{\omega}{6}}\,\tilde{k}\,\sqrt{\vert 1-\tilde{k}^2\vert}\Bigg\{
M_{GW}^2 (r-1)\nonumber\\
&&\qquad\qquad\qquad\qquad\qquad + H^2\,\left[2\,r\,(8\,r-\omega-10)+\omega+4 - \frac{(2\,r-1)(8\,r-\omega-7)}{(1-\tilde{k}^2)}-\frac{3\,(2\,r-1)}{(1-\tilde{k}^2)^2}\right]\Bigg\}\,,\nonumber\\
(\Omega^2)_{22} &=&
M_{GW}^2 + \frac{(6-\omega)\,\tilde{k}^2}{3}\,\Bigg\{
M_{GW}^2(r-1)+2\,H^2\,\left[r(4\,r-5)-\frac{\omega}{8}+1-\frac{(2\,r-1)^2}{1-\tilde{k}^2}
\right]\Bigg\}\,,
\label{qmg-freqm}
\end{eqnarray}
where $M_{GW}^2$ was defined in Eq.(\ref{mgwqmgdef}). For $\tilde{k}>1$, the mode $Z_1$ becomes a ghost, while for momenta $\tilde{k}<1$, both degrees of freedom are well-behaved.

At low momenta ($\tilde{k}\ll 1$), the action effectively becomes diagonal. In this long wavelength regime, both degrees of freedom have positive kinetic terms, while ${\cal M}_{ij}=0$ and the two eigenfrequencies are $\omega^2_1 = 0$ and $\omega^2_2 = M_{GW}^2$ \cite{Haghani:2013eya}. On the other hand, the ghost degree of freedom appears at momenta $\tilde{k} \gtrsim 1$, and so we must still diagonalize the system to determine the amplitudes of the frequencies in this regime.

\subsection{Diagonalization}
To find the eigenfrequencies of the system, first note that since the matrices ${\cal M}$ and $\Omega^2$ are time dependent, it is not possible to diagonalize the system at the level of the Lagrangian. On the other hand, the Hamiltonian can be written as a sum of decoupled oscillators, as shown in Ref.~\cite{Nilles:2001fg} and in the presence of ghosts, in Appendix D of Ref. \cite{DeFelice:2013awa}.
We consider the two cases ${\tilde{k} <1}$ and ${\tilde{k} >1}$ separately. 

\subsubsection{No ghost: ${\tilde{k} <1}$}

For the first case (denoted by subscript $<$), the kinetic matrix is unity and there is no ghost. In this case, we can introduce a rotated basis $ W \equiv R_<\,Z$, where the SO(2) rotation $R_<$ satisfies
\begin{equation}
\dot{R}_< = N\,R_<\,{\cal M}_<\,.
\end{equation}
This rotation allows us to remove the mixing, and the Lagrangian becomes
\begin{equation}
{\cal L}_< = \frac{\dot{W}^\dagger}{N} \,\frac{\dot{W}}{N} - W^\dagger\,R_<^T\,\left(\Omega^2_< + {\cal M}_<^T\,{\cal M}_< \right)R_<\,W \,.
\end{equation}
As shown in Appendix D of Ref.~\cite{DeFelice:2013awa}, the eigenvalues of the matrix 
\begin{equation}
\tilde{\Omega}^2_< \equiv \Omega^2_< + {\cal M}_<^T\,{\cal M}_<\,,
\end{equation}
correspond to the actual eigenfrequencies. The matrix $\tilde{\Omega}^2_<$ can be diagonalized by performing an SO(2) rotation $\xi_<$,
\begin{equation}
\xi^T_<\,\tilde{\Omega}^2_< \,\xi_< = \omega^2_<\,(\rm{diagonal})\,,
\end{equation}
where
\begin{equation}
\xi_< = \left(\begin{array}{cc}
               \cos(\theta_<) & \sin(\theta_<)\\
	      -\sin(\theta_<) & \cos(\theta_<)
              \end{array}
\right)\,,
\end{equation}
and
\begin{equation}
\sin( 2\,\theta_<) = \frac{(\tilde{\Omega}_<^2)_{12}}{\sqrt{\left[(\tilde{\Omega}_<^2)_{11}-(\tilde{\Omega}_<^2)_{22}\right]^2 +4\,\left[(\tilde{\Omega}_<^2)_{12}\right]^2}}\,,\qquad
\cos( 2\,\theta_<) = \frac{(\tilde{\Omega}_<^2)_{22}-(\tilde{\Omega}_<^2)_{11}}{\sqrt{\left[(\tilde{\Omega}_<^2)_{11}-(\tilde{\Omega}_<^2)_{22}\right]^2 +4\,\left[(\tilde{\Omega}_<^2)_{12}\right]^2}}\,.
\end{equation}
The eigenvalues are then
\begin{eqnarray}
(\omega^2_<)_1 &=& \frac{1}{2}\left[(\tilde{\Omega}_<^2)_{11}+(\tilde{\Omega}_<^2)_{22}-\sqrt{\left[(\tilde{\Omega}_<^2)_{11}-(\tilde{\Omega}_<^2)_{22}\right]^2 +4\,\left[(\tilde{\Omega}_<^2)_{12}\right]^2}\right]\,,\nonumber\\
(\omega^2_<)_2 &=& \frac{1}{2}\left[(\tilde{\Omega}_<^2)_{11}+(\tilde{\Omega}_<^2)_{22}+\sqrt{\left[(\tilde{\Omega}_<^2)_{11}-(\tilde{\Omega}_<^2)_{22}\right]^2 +4\,\left[(\tilde{\Omega}_<^2)_{12}\right]^2}\right]\,.
\end{eqnarray}

\subsubsection{One ghost: ${\tilde{k} >1}$} 

In this regime (denoted by subscript $>$), the kinetic matrix has Lorentzian signature and the first mode is a ghost. Again, we introduce a rotated basis $W \equiv R_>\,Z$, where the SO(1,1) rotation $R_>$ satisfies,
\begin{equation}
\dot{R}_> = N\,R_>\,{\cal M}_>\,\eta\,,
\end{equation}
with $\eta \equiv \rm{diag}(-1,1)$. This rotation removes the mixing, and the Lagrangian becomes
\begin{equation}
{\cal L}_> = \frac{\dot{W}^\dagger}{N} \,\eta\,\frac{\dot{W}}{N} - W^\dagger\,R_>^T\,\left(\Omega^2_> + {\cal M}_>^T\,\eta\,{\cal M}_> \right)R_>\,W \ .
\end{equation}
In this case, the matrix we need to diagonalize is \cite{DeFelice:2013awa}, 
\begin{equation}
\tilde{\Omega}^2_> \equiv \Omega^2_> + {\cal M}_>^T\,\eta\,{\cal M}_> \,,
\end{equation}
which gives the actual eigenfrequencies
\begin{equation}
\xi^T_>\,\tilde{\Omega}^2_> \,\xi_> = \eta\,\omega^2_>\,(\rm{diagonal})\,.
\end{equation}
For an SO(1,1) rotation given by
\begin{equation}
\xi_> = \left(\begin{array}{cc}
               \cosh(\theta_>) & \sinh(\theta_>)\\
	      \sinh(\theta_>) & \cosh(\theta_>)
              \end{array}
\right)\,,
\end{equation}
where
\begin{equation}
\sinh( 2\,\theta_>) = -\frac{2\,(\tilde{\Omega}_>^2)_{12}}{\sqrt{\left[(\tilde{\Omega}_>^2)_{11}+(\tilde{\Omega}_>^2)_{22}\right]^2 -4\,\left[(\tilde{\Omega}_>^2)_{12}\right]^2}}\,,\qquad
\cosh( 2\,\theta_>) = \frac{(\tilde{\Omega}_>^2)_{11}+(\tilde{\Omega}_>^2)_{2}}{\sqrt{\left[(\tilde{\Omega}_>^2)_{11}+(\tilde{\Omega}_>^2)_{22}\right]^2 -4\,\left[(\tilde{\Omega}_>^2)_{12}\right]^2}}\,,
\end{equation}
the eigenvalues are
\begin{eqnarray}
(\omega^2_>)_1 &=& \frac{1}{2}\left[(\tilde{\Omega}_>^2)_{22}-(\tilde{\Omega}_>^2)_{11}-\sqrt{\left[(\tilde{\Omega}_>^2)_{11}+(\tilde{\Omega}_>^2)_{22}\right]^2 -4\,\left[(\tilde{\Omega}_>^2)_{12}\right]^2}\right]\,,\nonumber\\
(\omega^2_>)_2 &=& \frac{1}{2}\left[(\tilde{\Omega}_>^2)_{22}-(\tilde{\Omega}_>^2)_{11}+\sqrt{\left[(\tilde{\Omega}_>^2)_{11}+(\tilde{\Omega}_>^2)_{22}\right]^2 -4\,\left[(\tilde{\Omega}_>^2)_{12}\right]^2}\right]\,.
\end{eqnarray}

\subsubsection{Combining the two regimes}
Now that we have the necessary tools to diagonalize the system for the two regimes of momenta, we unify the two results. We first note that from Eqs.(\ref{qmg-mixm}) and (\ref{qmg-freqm}), we have
\begin{equation}
(\Omega^2_<)_{11} =-(\Omega^2_>)_{11}\,,\qquad
(\Omega^2_<)_{22} =-(\Omega^2_>)_{22}\,,\qquad
\left[(\Omega^2_<)_{12}\right]^2 =-\left[(\Omega^2_>)_{12}\right]^2\,,\qquad
\left[({\cal M}_<)_{12}\right]^2 =-\left[({\cal M}_>)_{12}\right]^2\,,
\end{equation}
which imply
\begin{equation}
(\tilde{\Omega}^2_<)_{11} =-(\tilde{\Omega}^2_>)_{11}\,,\qquad
(\tilde{\Omega}^2_<)_{22} =-(\tilde{\Omega}^2_>)_{22}\,,\qquad
\left[(\tilde{\Omega}^2_<)_{12}\right]^2 =-\left[(\tilde{\Omega}^2_>)_{12}\right]^2\,,
\end{equation}
or
\begin{equation}
(\omega_<)^2_1 = (\omega_>)^2_1 \,,\qquad
(\omega_<)^2_2 = (\omega_>)^2_2 \,.
\end{equation}

Thus, it is straightforward to write down a unified expression for the dispersion relation, independent of the momentum regime of the modes. We obtain,
\begin{eqnarray}
\omega^2_{1,2} &=& \frac{\tilde{k}^2}{6}\,\Bigg\{
-H^2\left[\omega\left(8\,r^2+\tfrac{\omega}{2}-11(2\,r-1)\right) + 4\,\frac{(6-\omega)r^2+\omega(r-1)+3}{1-\tilde{k}^2}+\frac{9}{(1-\tilde{k}^2)^2}\right]+M_{GW}^2(6-\omega)(r-1)\Bigg\}\nonumber\\
&&+ \frac{M_{GW}^2}{2} \mp \frac{1}{2}\sqrt{{\cal A}^2 + \frac{2\,(6-\omega)\,\tilde{k}^2(1-\tilde{k}^2)}{3}{\cal B}^2}\,,
\end{eqnarray}
where a $-$ ($+$) sign corresponds to the first (second) eigenmode, and we have defined
\begin{eqnarray}
{\cal A} &\equiv& \frac{\tilde{k}^2}{3}\,\Bigg\{
-H^2\left[\omega(6-\omega)-2\,(r-1)[4\,(12-\omega)r-5\,\omega-12] + \frac{6+\omega+8(r-1)[r(6-\omega)-3]}{1-\tilde{k}^2}-\frac{9}{(1-\tilde{k}^2)^2}\right]\nonumber\\
&&\qquad \qquad \qquad\qquad\qquad\qquad\qquad\qquad\qquad \qquad\qquad \qquad\qquad \qquad\qquad \qquad
+M_{GW}^2(6-\omega)(r-1)\Bigg\} +M_{GW}^2\,,\nonumber\\
{\cal B} &\equiv& H^2 \left[2\,(r-1)\left[6-\omega+8(r-1)\right]-\omega-\frac{(8\,r-\omega-7)(2\,r-1)}{1-\tilde{k}^2}-\frac{3\,(2\,r-1)}{(1-\tilde{k}^2)^2}\right]+M_{GW}^2(r-1)\,.
\end{eqnarray}

\subsection{Stability}
\label{qmg-stability}
Since the modes with momenta $\tilde{k} >1$ are ghosts, we need to determine how serious this problem is. The ghost mode appears at (physical) momenta parametrically of the order of the Hubble rate. If the frequencies of these modes are larger than the UV cutoff of the theory, they are not within the regime of validity of the low energy effective theory and may be ignored. 

\begin{figure}
\includegraphics[width=7.5cm]{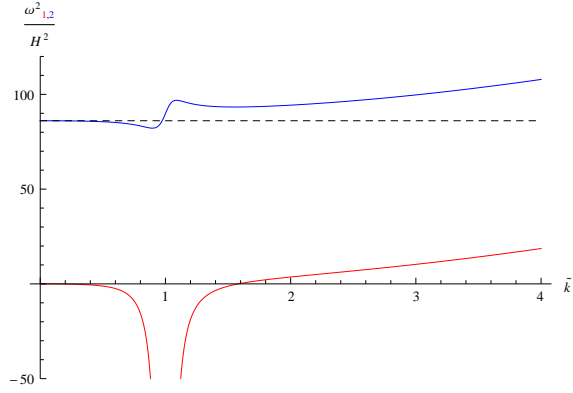}
\caption{Plot of the dispersion relation of scalar modes versus the rescaled momenta $\tilde{k}$, for the example (\ref{example}), with  $\Lambda=0$, $\omega=1$, $\alpha_3=-10$, $\alpha_4=6$, $\xi=0$, $m_g^2<0$, in the positive branch defined in Eq.(\ref{xpm}). The dashed line corresponds to $M_{GW}^2/H^2$ which is the mass term for mode 2 (blue) in the low momentum regime. At the critical point $\tilde{k}=1$, $\omega_1^2$ (red) diverges to $-\infty$, and then becomes positive and finite after the transition.
}
\label{noghostexample-sca}
\end{figure}

In the transition region where $\tilde{k} \to 1$, the frequencies are
\begin{eqnarray}
\omega_1^2 &=& -\frac{3\,H^2}{4\,(\tilde{k}-1)^2}+{\cal O}\left(\frac{1}{\tilde{k}-1}\right)\,,\nonumber\\
\omega_2^2 &=& M_{GW}^2+\frac{6-\omega}{3}\Bigg\{M_{GW}^2(r-1)+\frac{H^2}{6}\left[\omega(16\,r-5)+4\,r^2\left[2(6-\omega)\,r^2+4\,\omega\,r-6\,\omega-9\right]\right]\Bigg\}+{\cal O}(\tilde{k}-1) \,.
\end{eqnarray}
We note that the problematic mode, right after $\tilde{k}\sim1$, has a
very large frequency. As an example, we consider the set of parameters
\begin{equation}
\Lambda=0\,,
\qquad
\omega=1\,,
\qquad
\alpha_3 = -10\,,
\qquad
\alpha_4 = 6\,,
\qquad
\xi=0\,,
\qquad
m_g^2 <0\,,
\qquad
+\,{\rm branch}\,,
\label{example}
\end{equation}
where ``$+$ branch'' corresponds to the positive sign solution in Eq.(\ref{xpm}). These parameters lead to $r \simeq 1.01$ and $M_{GW}^2 \simeq 0.71\,\vert m_g^2\vert$, which satisfy the stability conditions for the tensor and vector modes. For this example, we show the momentum dependence of the scalar dispersion relations in Fig.\ref{noghostexample-sca}. As discussed in the paragraph after
Eq.(\ref{qmg-freqm}), at low momenta, $\omega_1^2\to 0$, while
$\omega_2^2 \to M_{GW}^2$. After the transition region, where
$\omega_1^2$ exhibits divergent behavior, both modes increase
with $\omega^2 \propto \tilde{k}^2$. The ``light'' mode, which becomes a
ghost in the $\tilde{k}>1$ region, has (for this specific example)
frequency $\omega_1/H \propto {\cal O}(1)$, so that apart from in the immediate 
neighborhood of $\tilde{k}\sim 1$, it cannot be integrated out from the
low energy effective theory. 
\begin{figure}
\includegraphics[width=6.5cm]{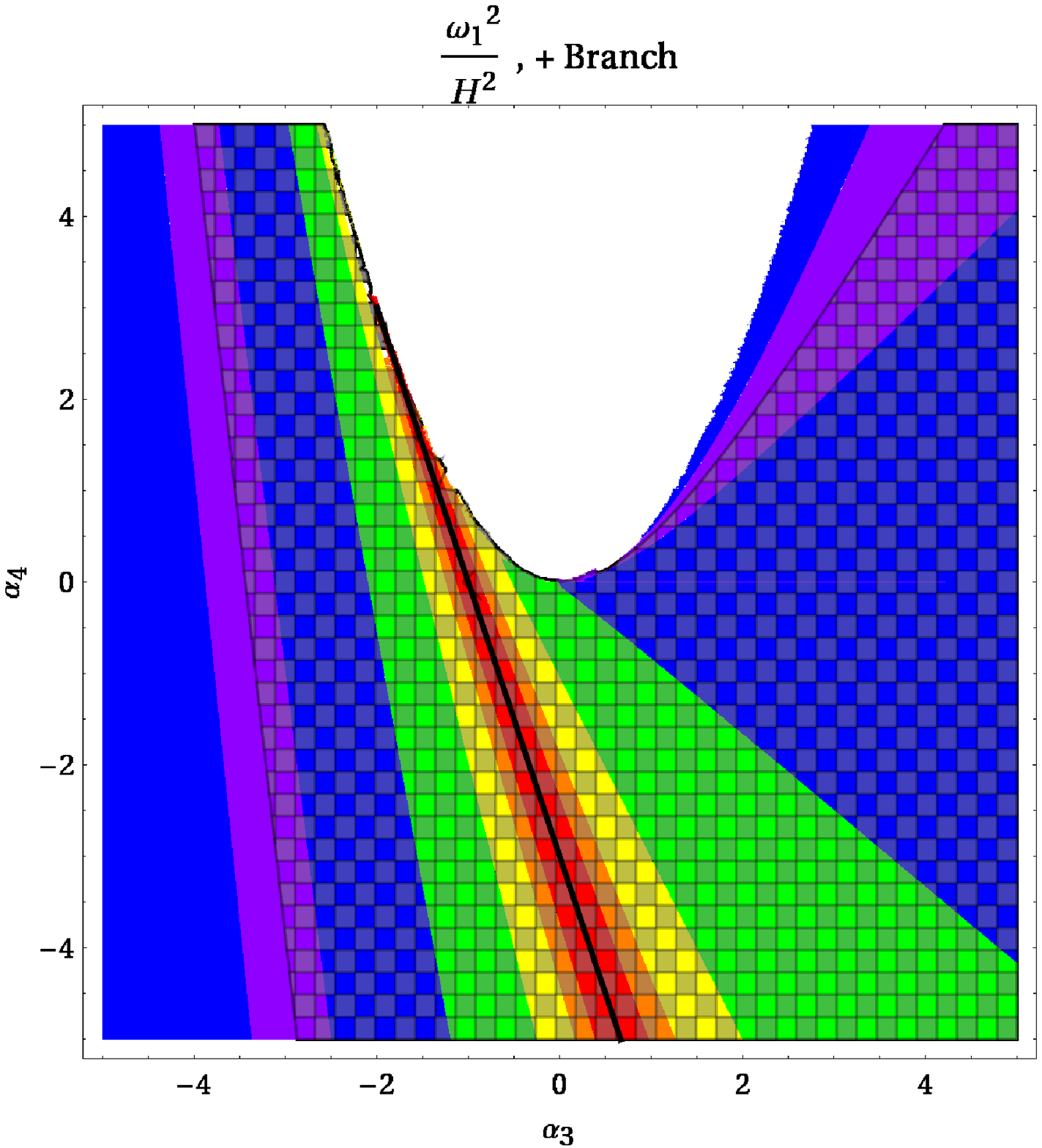}~~~~~~
\includegraphics[width=6.5cm]{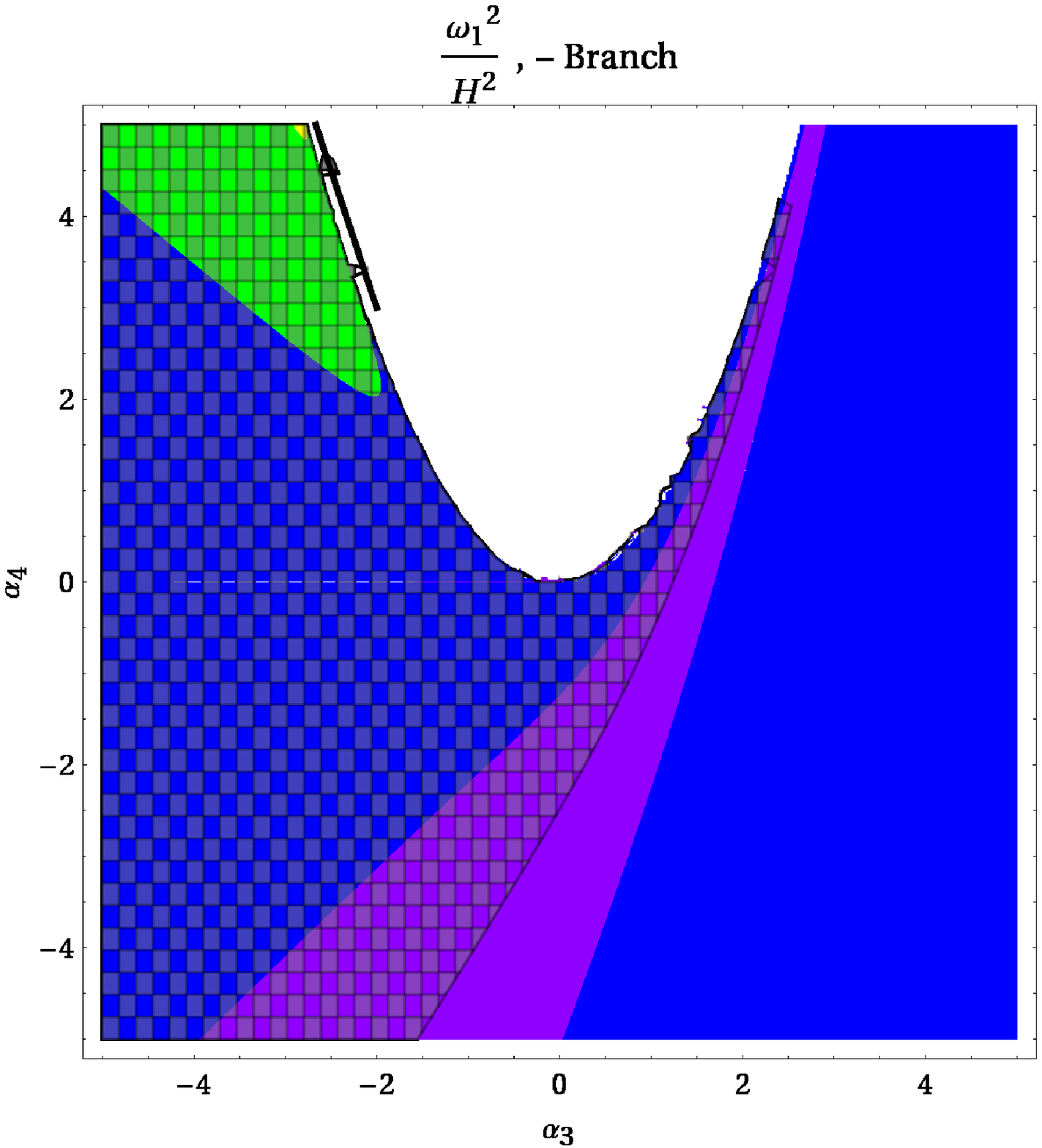}
\includegraphics[width=6.5cm]{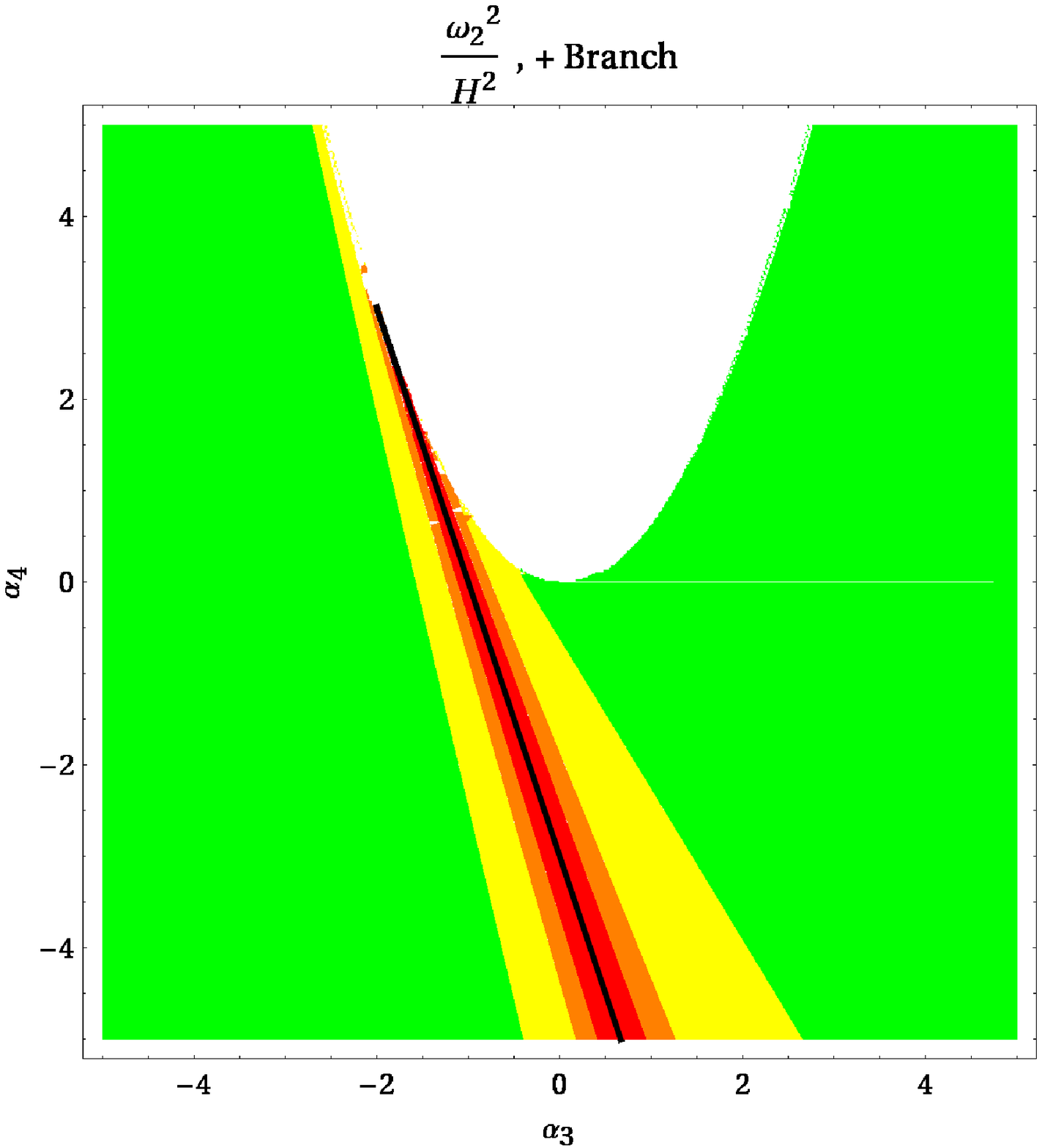}~~~~~~
\includegraphics[width=6.5cm]{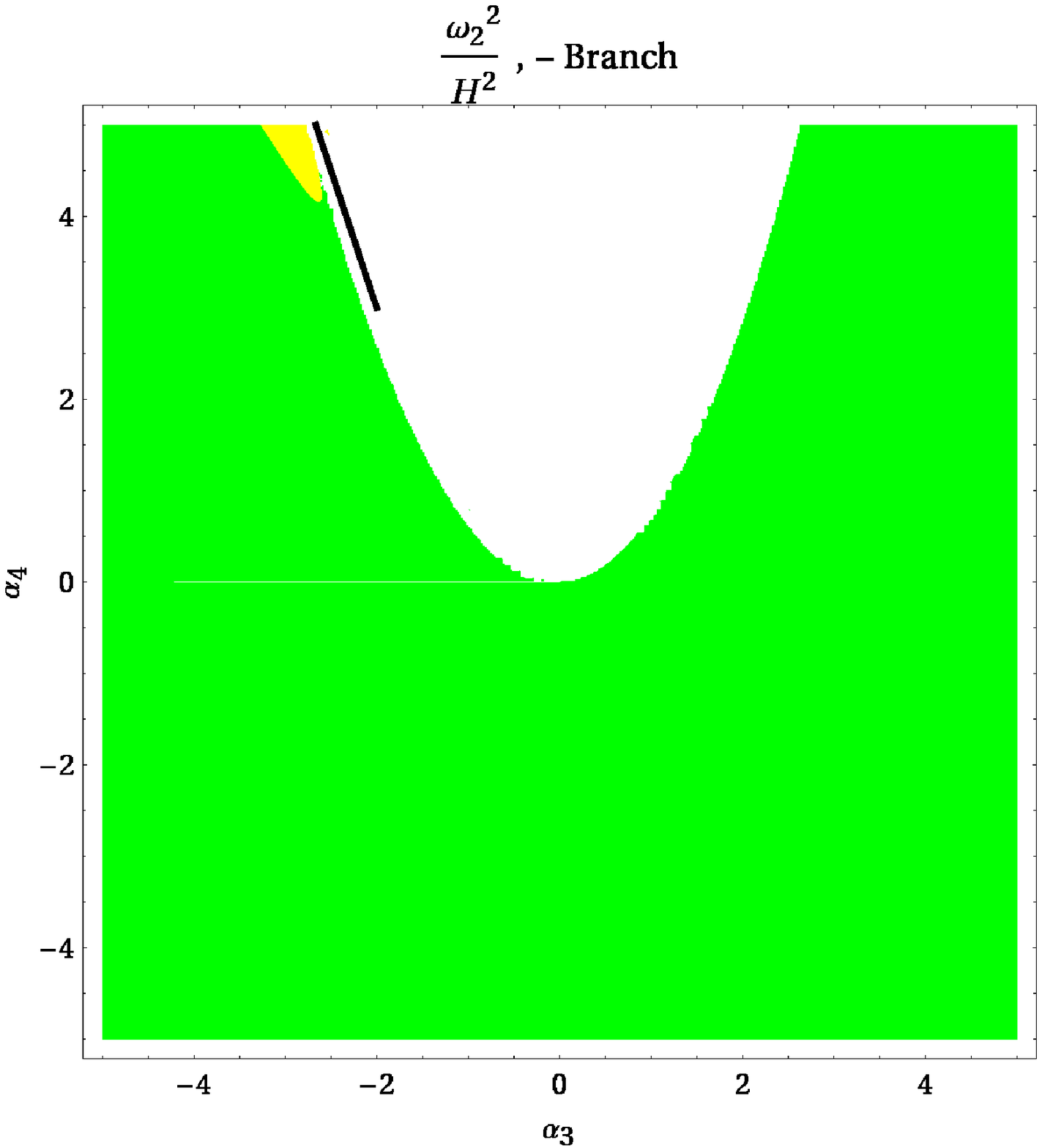}
\caption{Order of magnitude plots of the squared-frequency in Hubble units for the case $\Lambda=0$, $\omega=1$ and $\xi=0$, and for a benchmark momentum value $\tilde{k}=2$. The left (right) column shows the positive (negative) branch of solutions given in Eq.(\ref{xpm}), while the upper (lower) panel shows $\omega_1^2$ ($\omega_2^2$). The color representation is as follows: {\color[rgb]{.56,0,1} Violet - $(0,1]$}~; {\color{blue} Blue - $(1,10]$}~; {\color{green} Green - $(10,10^2]$}~; {\color[rgb]{.7,.7,0}  Yellow - $(10^2,10^3]$}~; {\color[rgb]{1,0.5,0} Orange- $(10^3,10^4]$}~; {\color{red} Red - $(10^4,\infty)$}~. The gridded region corresponds to $\omega^2 < 0$, while the thick black line corresponds to $\alpha_4 = -3(1+\alpha_3)$ [for the positive (negative) branch, $\alpha_3 >-2$ ($\alpha_3 <-2$) only], where $X\sim 0$.
}
\label{scaregdetail}
\end{figure}

Next, we consider a more general example, and extend our analysis to the $\alpha_3$, $\alpha_4$ parameter space. Instead of analyzing the immediate neighborhood of the critical point, we chose $\tilde{k}=2$, such that the frequency of the ghost mode becomes finite, while differing from the critical point value by an order one factor. In Fig.\ref{scaregdetail}, we show the order of magnitude of $\omega^2/H^2$ for an example with $\Lambda=0$,~$\omega=1$ and $\xi=0$, in both branches defined in (\ref{xpm}). In the regimes where $\omega_1^2>0$, the frequency is always of ${\cal O}(H)$, and the ghost mode cannot be removed. On the other hand, we see that in a special region, the ratio $\vert\omega_1\vert/H$ may exceed $10^2$. The source of large $\vert\omega_1\vert$ is related to a specific relation between parameters,
\begin{equation}
\alpha_4 = -3\,(1+\alpha_3)\,,\qquad
\left\{
\begin{array}{cc}
\alpha_3 >-2\,,& +{\rm ~branch} \\
\alpha_3 <-2\,,& -{\rm ~branch}
\end{array}
\right.\,,
\label{problemline}
\end{equation}
which leads to $X=0$ and  $r \propto X^{-2} \to \infty$. Although we have excluded the line (\ref{problemline}) from our analysis, parameters close to this line lead to the large (negative) values we observe in the squared-frequency of the ghost mode. In principle, the value of $\vert\omega_1\vert/H$ can be made arbitrarily large by tuning $\alpha_3$ and $\alpha_4$ to be close enough to this line. 

Finally, we note that in general, close to line (\ref{problemline}), the
parameter $M_{GW}^2$ defined in Eq.(\ref{mgwqmgdef}) becomes negative,
with $M_{GW}^2 = -\omega\,H^2 + {\cal O}(X)$. In other words, even if
the parameters are fine-tuned to remove the ghost mode in the scalar
sector, the tensor and vector modes have a tachyonic instability
(although its rate is at most of the Hubble scale). Additionally, the fact that $X\to0$ and $r\to \infty$ is also an indication of strong coupling, since the determinant of the kinetic matrix (\ref{detK}) is of the same small order of magnitude as the fine-tuning between $\alpha_4$ and $\alpha_3$.

\section{Comparison of the scalar sector of the quasi-dilaton theory with results in the decoupling limit}
\label{app:comparison}

In this Appendix, we compare our results for the scalar sector of the quasi-dilaton theory with those obtained in \cite{D'Amico:2012zv} in a decoupling limit. We first identify our degrees of freedom with the ones used by Ref.\cite{D'Amico:2012zv}. By comparing perturbations of $\delta g_{ij}$, we find that
\begin{equation}
\Psi = \frac{\partial^2b}{3\,m_g\,a^2}-\frac{H}{m_g}\,A_0\,,
\qquad
E=\frac{2}{m_g\,a^2}\,b\,,
\label{dghpfield}
\end{equation}
where $\Psi$ and $E$ are the perturbations introduced by the decomposition (\ref{eq:qmg-decomp}) and $A_0$, $b$ are the perturbations used in \cite{D'Amico:2012zv}. For the quasi-dilaton perturbations, $\delta\sigma$ in (\ref{eq:qmg-dilaton}) coincides with $\zeta$ in \cite{D'Amico:2012zv}.

We now turn to the action (\ref{eq:actionscaqmg}), which has the following kinetic term
\begin{equation}
S \ni \frac{M_p^2}{2}\int d^3k\,a^3\,N\,dt\,\frac{\dot{Y}^\dagger}{N} \,{\cal K}\,\frac{\dot{Y}}{N}\,,
\end{equation}
where the components of the kinetic matrix are given in Eq.(\ref{eq:qmg-scalarkin}) and the field basis is given by
\begin{equation}
Y \equiv \left( 
\begin{array}{c}
\frac{1}{\sqrt{2}\,k^2}\,\left(\Psi-\delta\sigma\right)\\\\ E
\end{array}
\right)\,,
\end{equation}
or, using Eq.(\ref{dghpfield}),
\begin{equation}
Y \equiv \left( 
\begin{array}{c}
\frac{1}{\sqrt{2}\,k^2}\left(\frac{\partial^2 b}{3\,m_g\,a^2}-\rho\right)
\\\\ \frac{2\,b}{m_g\,a^2}
\end{array}
\right)\,,
\end{equation}
with $\rho = \zeta+(H/m)A_0$.

Since the decoupling limit action in Ref.\cite{D'Amico:2012zv} is given in the $Z \equiv(\rho,b)$ basis (up to the non-dynamical degree of freedom $A_0$), we transform our action via
\begin{equation}
Y = R\,Z =
\left(
\begin{array}{cc}
 -\frac{1}{\sqrt{2}k^2} & -\frac{1}{3\,\sqrt{2}\,m_g\,a^2}\\\\
0 & \frac{2}{m_g\,a^2}
\end{array}
\right)
\left(
\begin{array}{l}
\rho\\\\
b
\end{array}
\right)\, ,
\end{equation}
so that the kinetic term in the $Z$ basis becomes
\begin{equation}
{\cal K}_{Z} = R^T\,{\cal K}\,R\,.
\end{equation}
Before taking the decoupling limit, we consider the determinant of the kinetic matrix, given by
\begin{equation}
\det {\cal K}_Z= \left(- \frac{\sqrt{2}}{k^2\,m_g\,a^2}\right)^2\,\det {\cal K}\,.
\label{dghpkin}
\end{equation}
In other words, the momentum dependent ghost-free condition (\ref{noghostsca}) is still valid. On the other hand, if we go to the decoupling limit, given by
\begin{equation}
m_g\to 0\,,\qquad
H\to0\,,\qquad
\frac{H}{m_g} = {\rm finite}\,,
\end{equation}
the kinetic matrix in the $Z$ basis becomes,
\begin{equation}
{\cal K}_Z= \left(
\begin{array}{cc}
\omega +{\cal O}(\epsilon^2) & {\cal O}(\epsilon)\\
{\cal O}(\epsilon)  & {\cal O}(\epsilon^2)
\end{array}
\right)\,,
\end{equation}
where $\epsilon$ denotes the order of $m_g$ and $H$. Thus, we see that at momenta comparable to and smaller than the expansion rate, one degree of freedom becomes a ghost, as we found in the main text. Therefore, this decoupling limit is not sufficient for determining the stability of one of the degrees of freedom. (Also see the discussion at the end of Sec.\ref{qdmgscalar}). This result coincides with the conclusion in \cite{D'Amico:2012zv}, from the determinant in Eq.(\ref{dghpkin}).

\section{Diagonal Basis for the scalar sector of the  varying mass theory}
\label{app:diagonalkin}
In this Appendix, we diagonalize the kinetic matrix of the scalar sector in the varying mass gravity theory, studied in Section \ref{vm-scalar}. Specifically, we want to show that the condition
\begin{equation}
\det{\cal K} > 0\,,
\end{equation}
for the kinetic matrix is enough to ensure the absence of ghost degrees of freedom.

The kinetic part of the action is given by
\begin{equation}
S \ni \int \frac{d^3k}{2}\,a^3\,N\,dt\,\frac{\dot{Y}^\dagger}{N} \,{\cal K}\,\frac{\dot{Y}}{N} \,,
\end{equation}
where the components of the kinetic matrix are given in Eq.(\ref{eq:kinmatvm}) and $Y = (\tilde{\delta\sigma},\,E)$. We now define a new basis,
\begin{equation}
Z_1 \equiv \frac{k^3\,M_p}{3\,a\,H}\,\left[E + 6\,\sqrt{2}\left(1- \frac{3\,a^2\,(\rho_m + p_m)}{2\,k^2M_p^2(r-1)}\right)\tilde{\delta\sigma}\right]\,,\qquad
Z_2 \equiv \frac{k^2\,M_p}{\sqrt{6}}\left(E + 6\,\sqrt{2}\,\tilde{\delta\sigma}\right)\,,
\end{equation}
after which, the kinetic terms become diagonal
\begin{equation}
S \ni \int \frac{d^3k}{2}\,a^3\,N\,dt\,\left(\kappa_1\,\frac{\dot{Z}_1^\dagger}{N} \,\frac{\dot{Z}_1}{N} + \kappa_2\,\frac{\dot{Z}_2^\dagger}{N} \,\frac{\dot{Z}_2}{N}\right)\,,
\end{equation}
with
\begin{equation}
\kappa_1 = \left[
\frac{k^2}{a^2\,H^2\,\left(\frac{\rho_\sigma+p_\sigma}{4\,M_p^2\,H^2}-\frac{3}{2}\right)}-\frac{\rho_m+p_m}{M_p^2\,H^2}
\right]^{-1}\,,
\qquad
\kappa_2 = 1\,.
\end{equation}
Thus, the condition (\ref{eq:vm-noghostsca2}), obtained from the positivity of $\det {\cal K}$, actually corresponds to the sign of the kinetic term of $Z_1$, while $Z_2$ always has positive kinetic term.

\end{document}